\documentclass[prd,superscriptaddress,showpacs]{revtex4}

\usepackage{epsfig}
\usepackage{graphicx}
\usepackage{latexsym}
\usepackage{amsmath}

\begin{document}

\title{High Energy Neutrino, Photon and Cosmic Ray Fluxes from VHS 
Cosmic Strings}
\date{\today}

\author{Ubi F. Wichoski}
\email{wichoski@ams3.ist.utl.pt}
\affiliation{Physics Department, Brown 
University, Providence, RI 02912, USA.}
\affiliation{Depto. de F\'{\i}sica, CENTRA-IST, 
Av. Rovisco Pais, 1 - Lisbon 1049-001, Portugal}

\author{Jane H. MacGibbon}
\email{jane.macgibbon1@jsc.nasa.gov}
\affiliation{Code SN3, NASA Johnson Space 
Center, Houston, TX 77058, USA}
  
\author{Robert H. Brandenberger}
\email{rhb@het.brown.edu} 
\affiliation{Physics Department, Brown 
University, Providence, RI 02912, USA.}

\begin{abstract}
Decaying topological defects, in particular cosmic strings, can 
produce a significant flux of high energy neutrinos, photons and 
cosmic rays. According to the prevailing understanding of cosmic 
string dynamics in an expanding Universe, the network of long 
strings loses its energy first into string loops, which in turn 
give off most of their energy as gravitational radiation. 
However, it has been suggested by Vincent \textit{et al.} 
(VHS) that particle emission may be the dominant energy 
loss channel for the long string network. In this case, 
the predicted flux of high energy particles would be much 
larger. Here we calculate the predicted flux of high 
energy gamma rays, neutrinos and cosmic ray antiprotons 
and protons as a function of the scale of symmetry breaking 
$\eta$ at which the strings are produced and as a function of the 
initial energy $m_J$ of the particle jets which result from the 
string decay. Assuming the validity of the VHS scenario, we find 
that due to the interactions with the cosmic radiation 
backgrounds all fluxes but the neutrino flux are suppressed at 
the highest energies. This indicates that the observed events 
above the GZK cutoff can only be accounted for in this scenario 
if the primary particle is a neutrino and $\eta$ is somewhat 
less than the GUT scale, i.e. $\eta \lesssim 10^{23}$ eV. The 
domain of parameter space corresponding to GUT-scale symmetry 
breaking is excluded also by the current observations below the 
GZK cutoff. A new aspect of this work is the 
calculation of the spectrum of the tau neutrinos
directly produced in the decay of the X-particles. This significantly 
increases the tau neutrino signal at high energies in all cosmic string 
scenarios.
\end{abstract}


\preprint{BROWN-HET-1115}

\maketitle


\section[intro]{Introduction} 
\label{sec:intro}

Measurements of the spectra of high energy $\gamma-$rays, neutrinos 
and cosmic rays have emerged as a useful constraint on particle physics 
theories which predict topological defects (for recent summaries 
see e.g. Refs. \cite{review,Sigl,Berez}). Although it does not 
appear likely that the emission of high energy particles by 
topological defects can explain the observed spectra, the current 
data can be used as upper bounds to constrain theories of particle 
physics beyond the Standard Model.

Of particular interest for cosmology are theories giving rise to 
cosmic strings. Some time ago, the spectrum of cosmic rays from 
non-superconducting strings was computed \cite{JR1,Pijush1,JR2} 
under the assumption of the \textit{standard} scaling 
picture \cite{Zel,Vil,Vil85} emerging from studies of cosmic 
string dynamics in an expanding Universe 
(see Refs. \cite{VSrev,HKrev,RHBrev} for recent reviews). In the 
standard picture, the long string network evolves into string 
loops which then decay predominantly by gravitational radiation. 
It was found that for strings with  $G \mu \sim 10^{-6}$ 
(required for strings to be relevant for cosmic structure 
formation \cite{TB86,Sato,Stebbins}), where $\mu$ is 
the mass per unit length in the string and $G$ is the Newton 
gravitational constant, the predicted particle 
fluxes are substantially lower than the observational detections 
or limits.

Recently, however, Vincent, Hindmarsh and 
Sakellariadou \cite{VHS} 
have challenged the standard picture of string evolution. 
They claim 
that the small-scale structure on the strings does not scale with 
the expansion of the Universe. In the \textit{VHS} scenario, the long 
strings lose their energy directly into particles instead of 
string loops. This leads to a greater production rate of particles 
and hence to the expectation of greater fluxes of $\gamma-$rays, 
neutrinos, and cosmic rays at Earth. 

Expanding on these points further, detailed numerical 
simulations \cite{AT,BB,AS} have demonstrated 
that in an expanding Universe, the network of \textit{long} 
cosmic strings (\textit{long} meaning with curvature radius 
greater than the Hubble radius) approaches a scaling 
solution in which the number of long string segments crossing 
each Hubble radius approaches a constant value $\nu$. 
In the VHS scenario, the string network maintains this scaling 
solution by the direct radiation of scalar and gauge particles. 
In the standard scenario, the string network maintains the 
scaling solution by continuously giving off part of its energy 
in the form of string loops with radius smaller than the Hubble 
radius. Based on certain theoretical arguments 
(see e.g. \cite{ACK}) it is expected that the distribution of loops 
in the standard scenario also eventually takes on a 
scaling solution, i.e. when all lengths are scaled to the Hubble 
radius, the statistical properties of the distribution of string 
loops are independent of time. However, the resolution of string network 
simulations (Nambu action strings) is not yet good enough to be 
able to verify the standard scenario. Moreover, recent field theory 
simulations \cite{VAH} provide some evidence that the string 
defects give off their energy not in the form of string loops, 
but directly by scalar and gauge particle radiation, thus 
corroborating the VHS scenario (These latter simulations 
remain controversial \cite{EPSS,OBP}).

In the standard cosmic string scenario, the string loops oscillate 
due to the relativistic tension, and decay slowly by the 
emission of gravitational radiation. Only a small fraction of the energy 
is released in the standard scenario in the form of scalar and 
gauge particle radiation. This occurs via the process of cusp 
annihilation \cite{RB87}, during the final loop collapse or by the 
evaporation of black holes created from string loops \cite{BHrem}. 
It has been shown \cite{JR1,Pijush1,JR2,JRU1} that for GUT-scale 
($G \mu \sim 10^{-6}$) strings the flux of ultra-high energy (UHE) 
particles in the standard picture from these mechanisms is at or 
below the relevant observational measurements or limits. The predicted 
flux increases, and may be observable, if the symmetry breaking scale is 
significantly lower. 

In the VHS scenario, in contrast, all of the string energy is 
released directly from the long strings as scalar and gauge particle 
radiation. Additionally, the predicted flux increases as the symmetry 
breaking scale increases. Hence, the flux of high energy 
$\gamma-$rays, neutrinos and cosmic rays in the VHS scenario 
is expected to be much larger than the fluxes from loops in the 
standard scenario. 
Indeed, it has been remarked \cite{review,VAH,Pijush98b} 
that the fluxes from GUT scale strings in the VHS scenario will 
be above the observational limits. In this paper, 
we point out that this limit comes only from the neutrino emission at 
high energy and the cascades produced by the electromagnetic emission, 
whilst the nearest expected VHS string is too distant to produce an 
observable flux of cosmic rays or $\gamma$-rays. In the VHS scenario 
the highest energy fluxes (above the GZK cutoff) of cosmic rays and 
$\gamma$-rays are suppressed due to the interaction of these 
particles with the 
photons of the cosmic radiation backgrounds. The result is that no 
isotropic flux of cosmic rays and $\gamma$-rays from a network of 
VHS strings is expected to be observed. However, the 
neutrino emission from low $G \mu$ VHS strings may produce an 
observable signal.
We present a detailed calculation in this paper of the neutrino, 
photon and cosmic ray fluxes for a range of string and decay scales. 
We also show that the predictions depend sensitively 
on the initial energy $m_J$ characterizing the decay of the 
particle radiation, which in turn depends on the presently unknown 
physics in the regime between the scale of electroweak symmetry 
breaking and the scale of string formation. We find that decreasing 
$m_J$ leads to a decrease in the upper cutoff in the predicted spectrum 
and an increase in the cosmic ray fluxes at intermediate 
energies. Thus, even the additional freedom of decreasing 
$m_J$ cannot make GUT strings obeying VHS dynamics consistent 
with the observational constraints. 
Nevertheless, in the event that a segment of long string is 
substantially closer than the expected average distance to the nearest 
string, a highly anisotropic flux of cosmic rays protons, antiprotons 
and $\gamma$-rays could be observed above the GZK cutoff in the 
VHS scenario. This possibility, however, is not supported by the 
current observations of events above the GZK cutoff 
which exhibit large scale isotropy and at most only small scale 
clustering in arrival angle \cite{agasanew}.

In this paper we perform for the first time the calculation of the tau 
neutrinos directly produced by the hadronic decays of the X-particles. The 
calculation of the direct $\nu_{\tau}$ spectrum has been not previously 
been included in the analysis of any cosmic string scenario. This was 
because it was assumed that the $\nu_{\tau}$ spectrum was highly suppressed 
at all energies compared with the $\nu_{e}$ and $\nu_{\mu}$ spectra. Instead 
only the much weaker $\nu_{\tau}$ spectrum generated by the cascade of the 
X-decay $\nu_{e}$ and $\nu_{\mu}$ off the cosmic relic neutrino background 
was described. Recently, however, MacGibbon, Wichoski, 
and Webber \cite{MWW,MW} have shown 
that, in hadronic jets at accelerator energies, the tau neutrinos are a 
significant fraction of the total neutrino 
spectrum at high $x$, where $x$ is the ratio of the neutrino energy to the 
jet energy. Extrapolating their fragmentation function parametrization 
for $\nu_{\tau}$ to UHE energies, we show here that, in the VHS 
and all other cosmic string scenarios, the directly produced tau neutrinos 
generate a signal at Earth comparable at the highest energies with 
the expected $\nu_{e}$ and $\nu_{\mu}$ signal. This 
flux is orders of magnitude greater than the contribution from the 
cascade $\nu_{\tau}$. One consequence of our result is that the 
detection of a significant fraction of $\nu_{\tau}$ in UHE neutrino 
events may be due to hadronic decay at the source and not 
$\nu_{\mu} \rightarrow \nu_{\tau}$ oscillation in transit.

In the following sections, we present the neutrino, photon and cosmic ray 
spectra for vanilla emission taking $m_{\nu} = 0$ and omitting SUSY 
particles in the jet decay. However, we discuss in Section \ref{sec:ext} 
the effects of 
including a SUSY sector, non-zero neutrino mass, neutrino oscillation 
or other extensions to the Standard Model. In the Appendix, we perform the 
full computation of the flux of secondary muon and electron neutrinos 
resulting from muon decay in the particle jets. We use units in which 
$\hbar=c=G=1$, unless otherwise stated, and take a Hubble constant 
of $h_0 = 0.65$.

\section[jetff]{Jet Formation and Fragmentation} 
\label{sec:jet}

If the network of long strings scales, then the corresponding 
energy density 
$\rho_{\infty}$ at time $t$ is
\begin{equation}
\rho_{\infty} \, = \, \nu \mu t^{-2} \, ,
\end{equation}
where $\mu$ is the mass per unit length in the string. 
This equation defines the constant $\nu$ which counts the number 
of strings per volume $t^3$ in the scaling solution. The value of 
$\nu$ can be determined from simulations of cosmic string network 
evolution. Current work \cite{AT,BB,AS} gives $\nu \simeq 13$.
In the VHS scenario, the scaling solution is maintained not by the 
production of string loops, but by the radiation of gauge and 
scalar particles (which we collectively call X-particles in the 
following). Making use of the equation of state of a string, the 
continuity equation becomes
\begin{equation}
{\dot \rho_{\infty}} + 2 H \rho_{\infty} \, = \, - m_X \frac{d n_X}{d t} \,
\end{equation}
where $n_X$ and $m_X$ are the number density and mass, 
respectively, of the X-particles. $m_X$ is effectively the symmetry 
breaking scale at which the string formed, 
\begin{equation} 
m_X \sim \eta \sim \mu^{1/2}
\end{equation}
(e.g. $m_X \sim 10^{16}$ GeV for $G \mu \sim 10^{-6}$). Strictly 
speaking, $\mu^{1/2}/\eta \sim O(1)$ with the exact factor depending 
on the form of the potential of the scalar field.

The decay of the high energy X-particles will lead to the 
production of jets analogous to the QCD jets seen in accelerators. 
To simplify the discussion we  assume that all jets in the X-decay 
have the same initial energy, $m_J$. In this case, the decay of 
a single X-particle will lead to $m_X / m_J$ jets, and the number 
density of jets generated by the energy release of long strings is
\begin{equation} \label{jetdensity}
\frac{d n_J}{dt} \, = \frac{2}{3} \frac{\nu \mu}{m_J} t^{-3} \, .
\end{equation}

The largest uncertainty in the calculation of the flux of high 
energy particles generated by strings comes from our ignorance 
of the structure of the jets at ultra-high initial energies.
Jet production has been studied in detail in QCD. From such 
studies, the fragmentation function of jets into photons, 
neutrinos and baryons is known \cite{Lund,Herwig}, at 
least to a good approximation, up to a few TeV. However, in our 
case the particles producing the jets are superheavy scalar and 
gauge particles. Following Refs. \cite{Hill83,HSW87}, the 
fragmentation functions of such jets are usually calculated by 
extrapolating the QCD fragmentation functions to higher energies 
(in which case $m_J$ is taken to be $m_X/2$). This is 
appropriate if there is no new physics which does not match 
this extrapolation between the electroweak symmetry breaking 
scale and the unification scale (the scale at which the defects 
are produced), but it is not justified if there is new physics 
which does not sufficiently match the 
extrapolation. We will parametrize ignorance 
about new physics and its effects on the jets from the  
X-particle decay by introducing a new scale $m_J$ (which may 
be the scale of the new physics) as the scale to which 
we can satisfactorily extrapolate the fragmentation functions 
from the QCD regime. We will then assume that the initial 
scalar or gauge X-particle produces $m_X / m_J$ jets with 
initial energy $m_J$ each. (In reality, there may be 
intermediate decay steps between the $m_X$ and $m_J$ 
decay scales.)
While this may or may not be
the precise description of the decay, it will at least 
help further quantify the dependence of the final fluxes on the 
uncertainties in the fragmentation process.

The initial jet particle decays into quarks and leptons on a 
time scale of ${\tilde \alpha} m_J^{-1}$, where ${\tilde \alpha}$ 
is the coupling constant associated with the physics at the 
energy scale $m_J$. The quarks then hadronize on a strong 
interaction time scale. More than 90\% of the total energy 
in a QCD jet goes into pions, with the majority of the remainder 
going into baryons (mainly neutrons, antineutrons, protons and 
antiprotons). On astrophysical time scales, 
the neutral pions decay into two photons, the charged pions 
decay into electron and muon neutrinos and electrons or positrons, 
and the neutrons decay into protons and leptons. 
In addition, a relatively small but still significant number of tau 
neutrinos are produced by the decay of heavy quarks and tau leptons. 
Non-jet leptons $l_{X}$ produced by the decay 
$X \longrightarrow q + l_{X}$ may also influence the observable 
$\gamma$-ray flux below $E \alt 10^{11}$ eV 
(see Section \ref{sec:obs}). 

The distribution of energies $E$ of the primary QCD jet 
decay products (predominantly equal numbers of 
$\pi^0, \pi^+$ and $\pi^-$) can be approximated by 
the fragmentation function \cite{HSW87}
\begin{equation} \label{primdecay}
\frac{d N'}{d x} \, \simeq \, \frac{15}{16} x^{-3/2} (1 - x)^2 \, , 
\end{equation}
where $x = E / m_J$ is the fraction of the jet energy carried by 
the decay product and $E$ continues down to $\sim 10^{9}$ eV. The two 
body decays of the primary $\pi^0, \pi^+$ or $\pi^-$ decay 
particles lead, on the average, to two photons and two primary 
muon neutrinos for every three pions. The energy distribution of 
the decay products can be obtained by integrating (\ref{primdecay}) 
from $\alpha x$ to $1$ with the invariant measure $dx / x$,
\begin{equation} \label{secdecay}
\frac{d N}{d x} \, \simeq \, \frac{5}{8} \alpha \bigl[\frac{16}{3} 
- 2 \alpha^{1/2} x^{1/2} - 4 \alpha ^{-1/2} x^{-1/2} + 
\frac{2}{3} \alpha^{-3/2} x^{-3/2} \bigr]\, ,
\end{equation}
where the constant $\alpha$ depends on the decay process 
being considered (see below). Eq. (\ref{secdecay}) 
applies to the final spectrum of photons produced in the 
jet. The photons result from $\pi^0$ decay. In this case, 
both of the decay particles have vanishing rest mass, and it 
is possible for a single photon to carry away the entire pion 
energy in the lab frame. Hence, $\alpha = 1$ for the photons. 

Electron and muon neutrinos result from the $\pi^+$ and $\pi^-$ decay. 
In the decay process $\pi^{\pm} \rightarrow \mu^{\pm} + 
\stackrel{\mbox{\tiny (-)}}{\nu_{\mu}}$, 
one of the decay products (the muon or antimuon) has nonvanishing 
rest mass. In this case, the integration limits when integrating 
(\ref{primdecay}) are not $x$ and $1$ (see e.g. \cite{Stecker} 
and \cite{Halzen}). The correction to the upper integration end 
for $\nu_e$ and $\nu_{\mu}$ is negligible, but not so the change 
in the lower integration limit for $\nu_{\mu}$ which becomes 
$\alpha x$ with
\begin{equation} \label{alph}
\alpha \, = \, \frac{m_{\pi}^2}{m_{\pi}^2 - m_{\mu}^2}  \, 
\simeq \, 2.34 \, ,
\end{equation}
when $E_{\pi} \gg m_{\pi} \simeq 140$ MeV, the 
charged pion mass, and $m_{\mu} \simeq 106$ MeV is the muon mass. 
Equation (\ref{secdecay}) describes the 
$\nu_{\mu}$ and ${\bar \nu}_{\mu}$ neutrinos 
immediately produced by the decay of the charged pions. 
The final spectrum of $\nu_{\mu}$ and ${\bar \nu}_{\mu}$ 
neutrinos, however, is comprised of the 
$\nu_{\mu}$ and ${\bar \nu}_{\mu}$ 
neutrinos produced by the subsequent decay of the $\mu^+$ and 
$\mu^-$, together with the contribution from those 
immediately produced by the decay of the $\pi^+$ and $\pi^-$. 
The final spectrum of $\nu_e$ and ${\bar \nu}_e$ neutrinos is 
produced by the decay of the muons and antimuons. Taking into 
consideration the spin polarization of the muons, we show in 
the Appendix that the final spectra of the $\nu_e$ and $\nu_{\mu}$ 
neutrinos generated by jet decay are approximately 
\begin{eqnarray}    \label{appres1}  
\frac{d N_{\nu_{\mu} + {\bar \nu_{\mu}}}}{d E_{\nu_{\mu}}} & \simeq &  
\frac{5}{8 m_J} \Bigg[ 12.48 
+ 0.44\;\biggl(\frac{E_{\nu_{\mu}}}{m_J} \biggr)^{-3/2} 
- 6.12\; \biggl(\frac{E_{\nu_{\mu}}}{m_J} \biggr)^{-1/2} 
\nonumber \\
& & {}- 7.16\; \biggl(\frac{E_{\nu_{\mu}}}{m_J} \biggr)^{1/2} \Bigg] \; , 
\qquad \quad E_{\nu_{\mu}} \le \frac{m_J}{\alpha} \; ,
\end{eqnarray}
for the muon neutrinos produced in the first stage of the pion decay 
\begin{eqnarray}   \label{appres1b}     
\frac{d N_{\nu_{\mu} + {\bar \nu_{\mu}}}}{d E_{\nu_{\mu}}}  & \simeq & 
\frac{5}{8 m_J} \Bigg[ 11.83 
+ 0.48\;\biggl(\frac{E_{\nu_{\mu}}}{m_J} \biggr)^{-3/2} 
- 5.80\; \biggl(\frac{E_{\nu_{\mu}}}{m_J} \biggr)^{-1/2} 
\nonumber \\
& & {}- 7.33\; \biggl(\frac{E_{\nu_{\mu}}}{m_J} \biggr)^{1/2} 
+ 0.97\; \biggl(\frac{E_{\nu_{\mu}}}{m_J} \biggr)^{2} 
\nonumber \\
& & {}- 0.15\; \biggl(\frac{E_{\nu_{\mu}}}{m_J} \biggr)^{3} 
\Bigg] \; , \qquad \quad E_{\nu_{\mu}} \le m_J \; , 
\end{eqnarray}
for the muon neutrinos produced by the decay of muons in the pion decay, and 
\begin{eqnarray}   \label{appres2}  
\frac{d N_{\nu_e + {\bar \nu_e}}}{d E_{\nu_e}} & \simeq &
\frac{5}{8 m_J} \Bigg[ 12.68 
+ 0.47\; \biggl(\frac{E_{\nu_{e}}}{m_J} \biggr)^{-3/2} 
- 5.96\;  \biggl(\frac{E_{\nu_{e}}}{m_J} \biggr)^{-1/2}
\nonumber \\
& & {}- 8.30\;  \biggl(\frac{E_{\nu_{e}}}{m_J} \biggr)^{1/2} 
- 0.11\; \biggl(\frac{E_{\nu_{\mu}}}{m_J} \biggr)^{1} 
\nonumber \\ 
& & {}+ 1.67\; \biggl(\frac{E_{\nu_{\mu}}}{m_J} \biggr)^{2} 
- 0.45\; \biggl(\frac{E_{\nu_{\mu}}}{m_J} \biggr)^{3} \Bigg] \; , \qquad 
\quad E_{\nu_{e}} \le m_J \; ,
\end{eqnarray}
for the final electron neutrinos. 

\begin{figure}[ht]
\begin{center}
   \mbox{\epsfig{figure=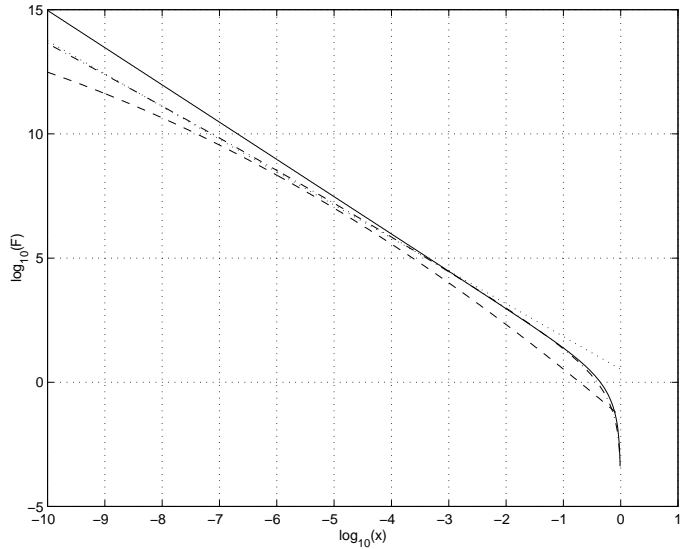,height=3.2in}}
\end{center}
\caption{Comparison of the three jet fragmentation functions 
discussed in the text. The solid line represents the fragmentation 
function (\ref{primdecay}) used in this paper to calculate all 
ultra-high energy particle fluxes but $\nu_{\tau}$, the dash-dotted 
line is the from of (\ref{ff2}), the dotted line is the approximation 
(\ref{ff2approx}) of (\ref{ff2}), and the dashed line is (\ref{ff3}).
\label{figfragfunc}}
\end{figure}

Equation (\ref{primdecay}) also applies to the primary baryonic 
decay products such as protons and neutrons. Assuming that 
3\% of the total energy of the jet goes into baryons and antibaryons 
and noting that most of the energy of decaying neutrons and 
antineutrons is transferred to daughter protons and antiprotons, 
the final distribution of protons and antiprotons is approximately
\begin{equation} \label{protonflux}
\frac{d N_{p + {\bar p}}}{d x} \, \simeq \, 0.03 \frac{15}{16} 
x^{-3/2} (1 - x)^2 \, .
\end{equation}

Equation (\ref{primdecay}) is one approximation to the 
numerically computed QCD fragmentation functions, taken 
from Ref. \cite{HSW87}. It is derived simply by requiring that 
the number of decay products scales as $E^{1/2}$ and applying 
energy conservation and scale invariance. Further the ratio of nucleon to 
pions is assumed at all $x$ to equal the ratio of total nucleons to total 
pions per jet. Even with these reservations, we find that Eqs. 
(\ref{appres1}) \textendash { }(\ref{appres2}) match the 
$\nu_e$ and $\nu_{\mu}$ spectra generated in 10 TeV 
$e^{+} e^{-} \rightarrow \; q \,\bar{q} \;$ events simulated by the 
HERWIG Monte Carlo code \cite{Herwig} to within a factor of 2 in the range 
$10^{-4} < x  < 1$ \cite{MW}.
There are other analytical approximations of the fragmentation functions 
used in the literature. Another form is \cite{Hill83,HSW87}
\begin{equation} \label{ff2}
\frac{d N'}{d x} \, \simeq \, 0.08 e^{2.6 \sqrt{ln(1/x)}} (1 -x)^2 
(x \sqrt{ln(1/x)})^{-1} \, ,
\end{equation}
which  can be well approximated by \cite{Pijush91}
\begin{equation} \label{ff2approx}
\frac{d N'}{d x} \, \simeq \, 3.322 x^{-1.324} 
\end{equation}
for values of $x$ between $10^{-10} < x < 10^{-2}$. 
A third formula is based on the Modified Leading Logarithmic 
Approximation (MLLA) \cite{Mueller} of QCD. At small 
$x$, the expression becomes \cite{Berez2,QCDbooks} 
\begin{equation} \label{ff3}
\frac{d N'}{d x} \, \simeq \, \frac{K}{x} 
exp \bigl[ - \frac{ln^2 x / x_m}{2 \sigma^2} \bigr] \, ,
\end{equation}
where 
\[ 
2 \sigma^2 = \frac{1}{6} ln \bigl( \frac{m_J}{\Lambda} 
\bigr)^{3/2} \;, 
\]
$x_m = (\Lambda/m_J)^{1/2}$ and $\Lambda = 0.234$ GeV. 
Conservation of total energy in the jet 
\[
\int_0^1 x \frac{d N'}{d x} d x = 1 
\]
gives the normalization constant $K$. In all approximations, 
extrapolations and Monte Carlo simulations of the fragmentation 
functions at high energies, the fraction of the jet energy carried 
by nucleons (or equivalently the nucleon multiplicity) lies in the 
range 3\textendash 10\% but the precise value and dependence on $x$ 
and $m_J$ is unknown \cite{sarkar}.

Fig. \ref{figfragfunc} shows a comparison of the three fragmentation 
functions (\ref{primdecay}), (\ref{ff2}) and (\ref{ff3}). The difference 
between these functions is small in the energy range 
$10^{-6} < x < 10^{-0.5}$, and will be unimportant for order of 
magnitude considerations. Hence, in the following we will use 
mostly approximation (\ref{primdecay}). We stress however that 
no extrapolated fragmentation function should be regarded as 
accurate to more than an order of magnitude at highest and lowest 
$x$. Investment in precision in those regions of the predicted spectra 
is unjustified and misleading.

For the first time in cosmic string emission analysis, we also 
include the distribution for the tau neutrinos produced in the jet decays, 
which had erroneously been assumed to be negligible in all previous work. 
While the number of $\nu_{\tau}$ produced per $m_J > 1$ TeV jet is 
less than $10^{-3}$ of the 
number of $\nu_{e}$ and $\nu_{\mu}$, the high energy $\nu_{\tau}$ are 
predominantly produced by the initial decays of the heavier quarks 
with shorter lifetimes and the 
$\nu_{e}$ and $\nu_{\mu}$ are produced by the final state cluster 
decays of the much lighter pions. This leads to significantly greater 
relative contribution from $\nu_{\tau}$ at high $x$ than previously 
assumed. The fragmentation distribution (\ref{primdecay}) is no longer 
relevant for the tau neutrino. Instead MacGibbon, Wichoski, and 
Webber \cite{MWW,MW} find 
that the fragmentation function for $\nu_{\tau}$  production 
in 300 GeV - 100 TeV $e^{+} e^{-} \rightarrow \; q \,\bar{q} \;$ events, as 
simulated by HERWIG \cite{Herwig} can be parametrized as

\begin{equation} \label{nutaufrag}
\frac{d N_{\nu_{\tau} + {\bar \nu_{\tau}}}}{d E_{\nu_{\tau}}}  \, \simeq \,
\frac{1}{m_J} \Bigg[ - 0.36 + 0.15\; \biggl(\frac{E_{\nu_{\tau}}}{m_J} 
\biggr)^{-1/2} 
+ 0.27\; \biggl(\frac{E_{\nu_{\tau}}}{m_J} \biggr)^{1/2} 
- 0.06\;\biggl(\frac{E_{\nu_{\tau}}}{m_J} \biggr)^{3/2} 
\Bigg] \; . 
\end{equation}
Consistent results are obtained from 
jet decay events simulated by  PYTHIA/JETSET \cite{Lund}.

The expressions for the number density of jets (\ref{jetdensity}), 
and for the energy distribution of the jet decay products, 
(\ref{secdecay}), (\ref{appres1}), (\ref{appres1b}), (\ref{appres2}), 
(\ref{protonflux}) or (\ref{nutaufrag}), can be convolved to obtain the 
expected  flux $F(E)$ of high energy photons, neutrinos, and cosmic 
rays with energy $E$ from a VHS string distribution 
\begin{equation} \label{flux}
F(E) = \frac{1}{4 \pi}\int_{t_c}^{t_{istr}} dt' \, 
\frac{d n_J}{d t'}\, \big(z(t') + 1\big)^{-3} \, \frac{d N}{d E'} \, 
\frac{d E'[E,z(t')]}{d E} \; .
\end{equation}

In Eq. (\ref{flux}), particles observed today ($t = t_0$) with energy $E$ 
were produced at a time $t'$ with energy $E' = E'[E,z(t')]$, where 
$z(t')$ denotes the redshift at time $t'$. The factor 
$(z(t') + 1)^{-3}$ expresses the dilution of the particle number 
density in an expanding Universe. The lower cutoff time $t_c(E)$ 
corresponds to the maximal redshift from which particles of 
present-day energy $E$ can reach us. This cutoff can be due 
either to interactions with the ambient extragalactic and galactic 
media during propagation, or to the constraint that the initial 
energy $E'$ must be less than the initial jet energy $m_J$. 
The upper cutoff time $t_{istr}$ corresponds to the 
latest time that the emission from the VHS cosmic string network 
can be considered isotropic (see Section \ref{sec:obs}). 
Thus our upper cutoff time is that from which emission will have just 
isotropized by today.

For protons, in order to calculate the Jacobian 
$\frac{d E'[E,z(t)]}{d E}$ 
we apply the continuous energy loss approximation (CEL) and solve 
Eq. (\ref{flux}) numerically.
The use of the CEL approximation is acceptable in this case where the 
distance to the sources is much larger than the attenuation length 
for the particle in the cosmic radiation backgrounds 
(see Section \ref{sec:obs}). 

For photons, if electromagnetic cascades are not taken yet into 
account (see below), we obtain (see Eq. (\ref{secdecay}) with 
$\alpha = 1$)
\begin{eqnarray} \label{fluxsec}
E^3 F_{\gamma}(E) & \simeq & \frac{1}{4 \pi} \, \frac{5}{2} 
\nu \mu t_0^{-2} m_J \Bigl(\frac{E}{m_J}\Bigr)^{3} 
\Biggl[ \frac{4}{3} ((z_{max} + 1) - (z_{min} + 1)) 
\nonumber \\
& & {}-  \frac{1}{3} ((z_{max} + 1)^{- 1/2} - (z_{min} + 1)^{- 1/2})
\Bigl(\frac{E}{m_J}\Bigr)^{- 3/2} 
\nonumber \\
& & {}- 2 ((z_{max} + 1)^{1/2} - (z_{min} + 1)^{1/2}) 
\Bigl(\frac{E}{m_J}\Bigr)^{- 1/2} 
\nonumber \\ 
& & {}- \, \frac{1}{3} ((z_{max} + 1)^{3/2} - (z_{min} + 1)^{3/2}) 
\Bigl(\frac{E}{m_J}\Bigr)^{1/2}  \Biggr] \, ,
\end{eqnarray} 

For neutrinos, if cascade off the relic neutrino background is not 
included, we have (see Eq. (\ref{appres2})) 
\begin{eqnarray} \label{nueflux}
E^3 F_{\nu_{e} + \bar{\nu}_{e}}(E) & \simeq &   
\frac{1}{4 \pi} \, \frac{5}{2} 
\nu \mu t_0^{-2} m_J \Bigl(\frac{E}{m_J}\Bigr)^{3} 
\Biggl[ {3.17}\, ((z_{max} + 1) - (z_{min} + 1))
\nonumber \\ 
& & {}- {0.24}\, ((z_{max} + 1)^{- 1/2} -  (z_{min} + 1)^{- 1/2}) 
\Bigl(\frac{E}{m_J}\Bigr)^{- 3/2}  
\nonumber \\
& & {}- {2.98}\, ((z_{max} + 1)^{1/2} - (z_{min} + 1)^{1/2}) 
\Bigl(\frac{E}{m_J}\Bigr)^{- 1/2} 
\nonumber \\
& & {}- {1.38}\, ((z_{max} + 1)^{3/2} - 
(z_{min} + 1)^{3/2}) \Bigl(\frac{E}{m_J}\Bigr)^{1/2} 
\nonumber \\
& & {}- {0.01} ((z_{max} + 1)^{2} - (z_{min} + 1)^{2}) 
\Bigl(\frac{E}{m_J}\Bigr) \nonumber \\
& & {}+ {0.14}\, ((z_{max} + 1)^{3} - (z_{min} + 1)^{3}) 
\Bigl(\frac{E}{m_J}\Bigr)^{2}  
\nonumber \\ 
& & {}- {0.03}\, ((z_{max} + 1)^{4} - (z_{min} + 1)^{4}) 
\Bigl(\frac{E}{m_J}\Bigr)^{3}
\Biggr]   \, ,
\end{eqnarray} 
for electron neutrinos and (see Eqs. (\ref{appres1}) and 
(\ref{appres1b}))

\begin{equation} \label{numuflux}
E^3 F_{\nu_{\mu} + \bar{\nu}_{\mu}}(E) \simeq \left\{
\begin{array}{l}
\frac{1}{4 \pi} \, \frac{5}{2} 
\nu \mu t_0^{-2} m_J \Bigl(\frac{E}{m_J}\Bigr)^{3} 
\Biggl[ {6.08}\, ((z_{max} + 1) - (z_{min} + 1)) \\ 
- \, {0.46}\, ((z_{max} + 1)^{- 1/2} 
\,- \, (z_{min} + 1)^{- 1/2}) \Bigl(\frac{E}{m_J}\Bigr)^{- 3/2} \\
- \, {5.96}\, ((z_{max} + 1)^{1/2} - 1) 
\Bigl(\frac{E}{m_J}\Bigr)^{- 1/2} \\
- \, {2.42}\, ((z_{max} + 1)^{3/2} 
\, - \, (z_{min} + 1)^{3/2}) \Bigl(\frac{E}{m_J}\Bigr)^{1/2} \\
+ \, {0.08}\, ((z_{max} + 1)^{3} - (z_{min} + 1)^{3}) 
\Bigl(\frac{E}{m_J}\Bigr)^{2} \\ 
- \, {0.01}\, ((z_{max} + 1)^{4} - (z_{min} + 1)^{4}) 
\Bigl(\frac{E}{m_J}\Bigr)^{3}
\Biggr]    \\ 
\mbox{for} \quad E \le \frac{m_J}{\alpha} \; ,
\\ \\
\frac{1}{4 \pi} \, \frac{5}{2} 
\nu \mu t_0^{-2} m_J \Bigl(\frac{E}{m_J}\Bigr)^{3} 
\Biggl[ {2.96}\, ((z_{max} + 1) - (z_{min} + 1)) \\
- \, {0.24}\, ((z_{max} + 1)^{- 1/2} - 
(z_{min} + 1)^{- 1/2}) \Bigl(\frac{E}{m_J}\Bigr)^{- 3/2} \\ 
- \, {2.90}\, ((z_{max} + 1)^{1/2} - (z_{min} + 1)^{1/2}) 
\Bigl(\frac{E}{m_J}\Bigr)^{- 1/2} \\
- \, {1.22}\, ((z_{max} + 1)^{3/2} - (z_{min} + 1)^{3/2}) 
\Bigl(\frac{E}{m_J}\Bigr)^{1/2} \\
+ \, {0.08}\, ((z_{max} + 1)^{3} - (z_{min} + 1)^{3}) 
\Bigl(\frac{E}{m_J}\Bigr)^{2} \\
- \, {0.01}\, ((z_{max} + 1)^{4} - (z_{min} + 1)^{4}) 
\Bigl(\frac{E}{m_J}\Bigr)^{3}
\Biggr]   \\
\mbox{for} \quad \frac{m_J}{\alpha} < E \le m_J \; ,
\end{array} \right.
\end{equation} 
for muon neutrinos. 

Similarly for the tau neutrinos, the flux from the 
string distribution neglecting cascading is (see Eq. (\ref{nutaufrag}))
\begin{eqnarray} \label{nutauflux}
E^3 F_{\nu_{\tau} + \bar{\nu}_{\tau}}(E) & \simeq & 
\frac{1}{4 \pi} \, 
\nu \mu t_0^{-2} m_J \Bigl(\frac{E}{m_J}\Bigr)^{3} 
\Biggl[ {-0.36}\, ((z_{max} + 1) - (z_{min} + 1)) 
\nonumber \\
& & {}+ {0.30}\, ((z_{max} + 1)^{1/2} -  (z_{min} + 1)^{1/2}) 
\Bigl(\frac{E}{m_J}\Bigr)^{- 1/2}  
\nonumber \\
& & {}+  {0.18}\, ((z_{max} + 1)^{3/2} - (z_{min} + 1)^{3/2}) 
\Bigl(\frac{E}{m_J}\Bigr)^{1/2} 
\nonumber \\
& & {}- {0.02}\, ((z_{max} + 1)^{5/2} - 
(z_{min} + 1)^{5/2}) \Bigl(\frac{E}{m_J}\Bigr)^{3/2} 
\Biggr]   \, ,
\end{eqnarray} 

In Eqs. (\ref{fluxsec}\textendash \ref{nutauflux}), $z_{min}$ is the 
redshift corresponding to $t_{istr}$ and $z_{max}(E)$ is 
the redshift corresponding to $t_c$:
\begin{equation} \label{zcutoff}
z_{max}(E) + 1 \, = \, min (z_{co}(E) + 1, \frac{m_J}{E}) \, ,
\end{equation} 
where $z_{co}(E)$ denotes the redshift cutoff due to astrophysical 
interactions.
For convenience, we have presented Eqs. 
(\ref{fluxsec} \textendash { }\ref{nutauflux}) 
for the case $z_{max}(E) < z_{eq}$. In our figures, we have 
calculated the more general result for $z_{max}(E) > z_{eq}$ 
where appropriate. 

The first important conclusion to draw from Eqs. 
(\ref{fluxsec}\textendash \ref{numuflux}) 
is that for energies $E$ substantially smaller than the effective 
initial jet energy $m_J$, the fragmentation term proportional to 
$(E / m_J)^{-3/2}$ similarly dominates the flux of photons, 
electron and muon neutrinos giving 
\begin{equation} \label{asymptemu}
E^3 F_{\gamma, \nu_{e} + {\bar \nu_{e}}, \nu_{\mu} + 
{\bar \nu_{\mu}}}(E) \, \propto \, m_J^{-1/2} \, 
\end{equation}
In the case of the tau neutrinos the 
fragmentation term proportional to $(E / m_J)^{-1/2}$ dominates the 
flux so that Eq. (\ref{nutauflux}) gives, 
\begin{equation} \label{asympttau}
E^3 F_{\nu_{\tau} + {\bar \nu_{\tau}}} \, \propto \, m_J^{-3/2} \, .
\end{equation}

Hence, for fixed cosmic string mass per unit length $\mu$, 
the spectrum of high energy photons and neutrinos 
increases as the jet energy $m_J$ decreases. This occurs because for 
smaller values of $m_J$ there are more jets emitted from the string 
and the fragmentation function as a function of $E$ is more compressed. 
These two effects overcome the $m_J^{3/2}$ or $m_J^{1/2}$ factor of the 
dominant term of the fragmentation function. We also note that the 
fluxes in the VHS scenario increase as $G \mu$ increases, or 
equivalently the symmetry breaking scale increases, in contrast to the 
general decrease seen in the standard cosmic string scenario. 

To complete the calculation of the photon, neutrino and cosmic 
rays  we need to know the values of $z_{max}(E)$. 
For protons and antiprotons with energy between 
$\sim 5 \times 10^{18}$ eV and $\sim 5 \times 10^{19}$ eV, the 
dominant interaction is 
$e^+ e^-$ pair production off the cosmic 
microwave background. At higher energies, photopion production 
becomes the dominant energy loss mechanism. The onset of 
photopion production corresponds to the GZK cutoff 
energy \cite{GZK}. Above this energy, the rate of energy 
loss increases dramatically, and nucleons do not reach us from 
cosmological distances.

\begin{figure}[ht]
\begin{center}
   \mbox{\epsfig{figure=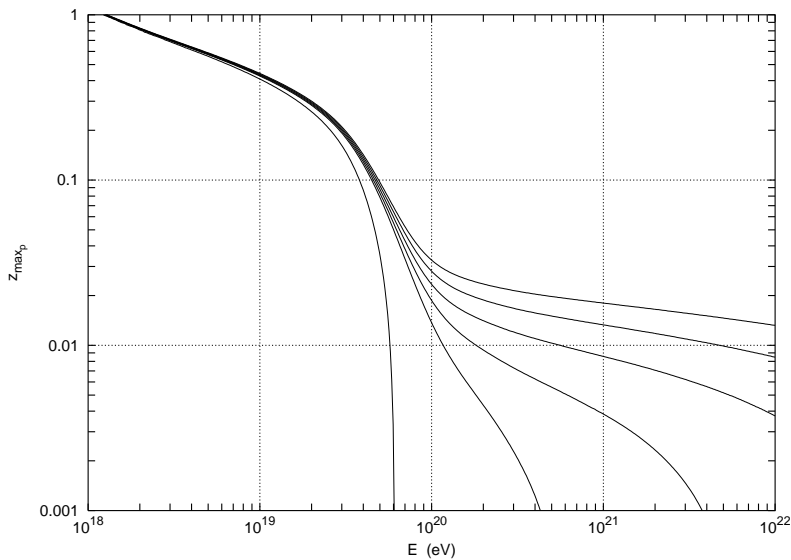,height=4.2in,angle=-90}}
\end{center}
\caption{The average maximal redshift $z_{max_p}(E)$ from which 
protons and antiprotons can reach the Earth as a function of the 
energy at arrival. 
The solid lines represent from top to bottom 
$G \mu = 10^{-6}, 10^{-8}, 10^{-10}, 10^{-12}, 10^{-14}, 10^{-16}$. 
Note the sharp decrease of the curve above 
$\sim 4 \times 10^{19}$ eV corresponding to the GZK cutoff.
\label{figpzmax}}
\end{figure}

Applying the results of \cite{absorp}, we find, in the case of 
protons and antiprotons, $z_{co_p}(E)$ for 
various values of $m_{J} = \eta/2$ (where $\eta$ is the scale 
of the symmetry-breaking) corresponding to 
$G \mu = 10^{-6}, 10^{-8}, 10^{-10}, 10^{-12}, 10^{-14},$ and 
$10^{-16}$ as  shown in Fig. \ref{figpzmax}. 

\begin{figure}[ht]
\begin{center}
   \mbox{\epsfig{figure=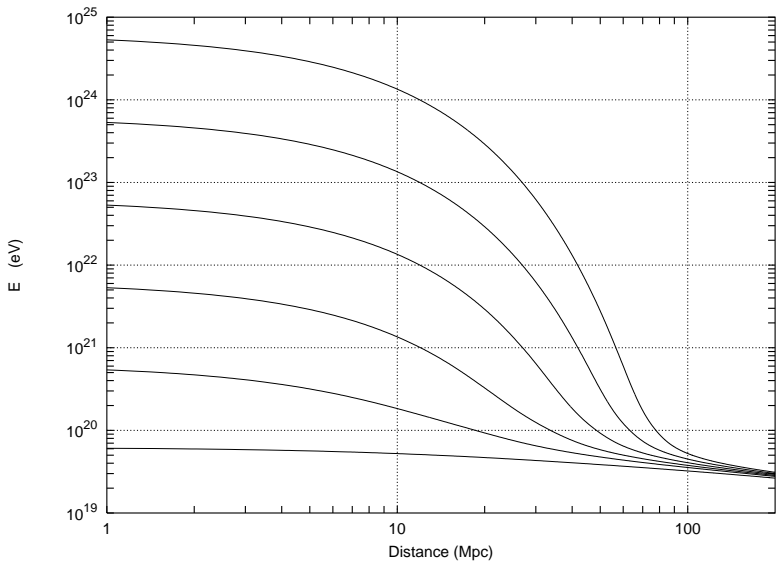,height=4.2in,angle=-90}}
\end{center}
\caption[...]{Energy of a proton (antiproton) as a function of the 
distance of travel through the CMB, for the same 6 values of $G \mu$ 
(determining the initial energy) used in the previous figure.
\label{figavdist}}
\end{figure}

In Fig. \ref{figavdist} we plot the average maximum distance that a 
ultra-high energy proton or antiproton can travel in the CMB.  
After propagating $\sim 100$ Mpc the proton or antiproton energy decreases 
below $\sim 6 \times 10^{19}$ eV irrespective of the initial jet energy.
For an expanded treatment see e.g. \cite{prev}.  

\begin{figure}[ht]
\begin{center}
   \mbox{\epsfig{figure=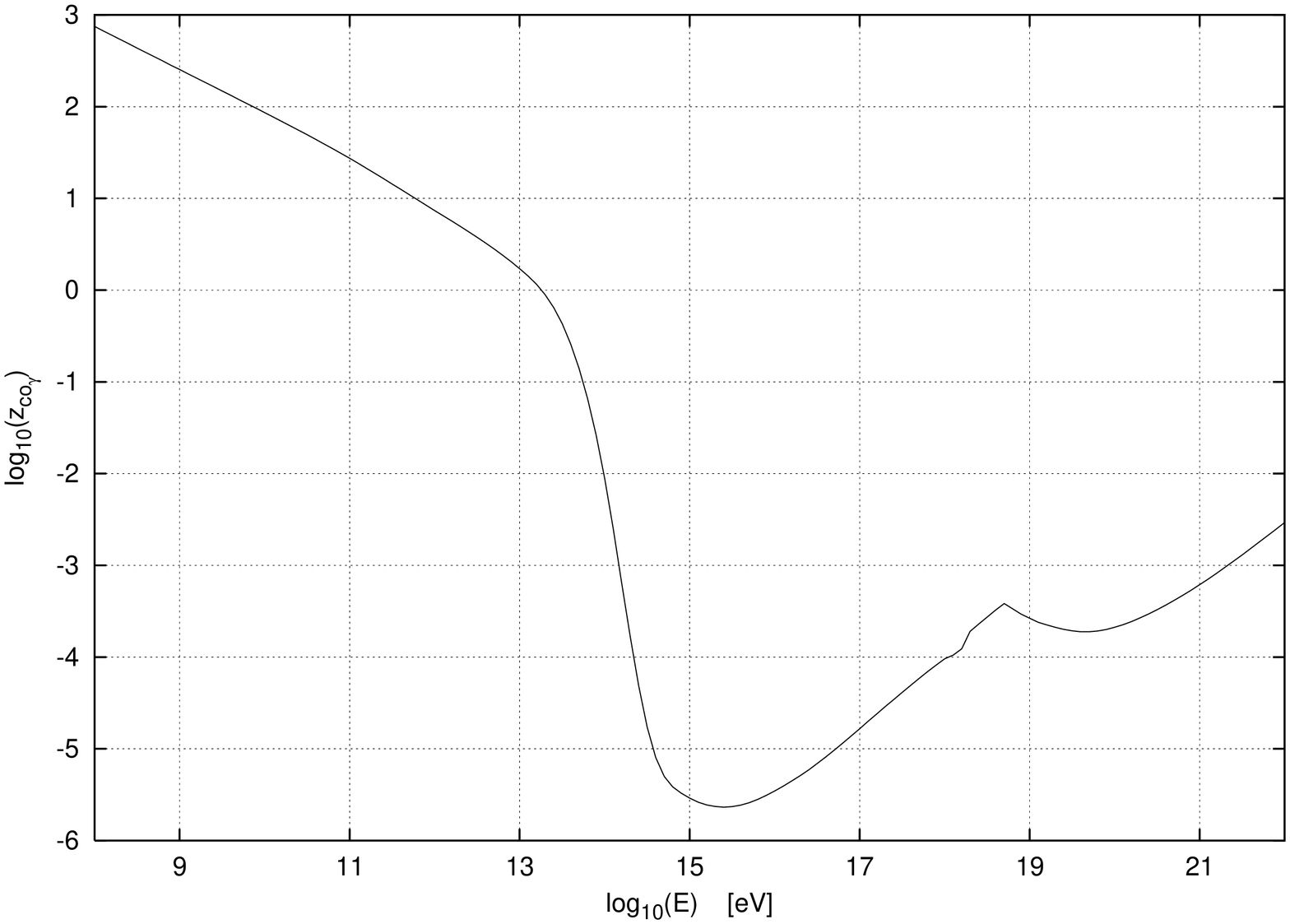,height=4.2in,angle=-90}}
\end{center}
\caption[...]{The maximal redshift $z_{co_{\gamma}}(E)$  from which 
photons can reach the Earth as a function of the energy at 
arrival \cite{JR2}.
\label{figphotabs}} 
\end{figure}

For photons of energies greater than about $10^{11}$ eV, 
$z_{co_{\gamma}}(E)$ is determined by pair production off the cosmic 
photon background. 
The resulting function $z_{co_{\gamma}}(E)$ is shown in 
Fig. \ref{figphotabs} 
(taken from Ref. \cite{JR2}). For lower energies, pair production off 
nuclei determines the form of $z_{co_{\gamma}}$. 

For neutrinos, the dominant process is scattering off the relic 
$1.9^o$K cosmic background neutrinos. In this case, $z_{co_{\nu}}(E)$ 
is given by \cite{BHS92} 
\begin{equation} \label{zn1}
z_{co_{\nu}} \, \simeq \, 9 \times 10^{8} \bigg(\frac{E}{eV}\bigg)^{-1/3} 
\, , \quad E \,\leq \, 3 \times 10^{14} \; eV \,
\end{equation}
and
\begin{equation} \label{zn2}
z_{co_{\nu}} \, \simeq \, 5 \times 10^{8} \bigg(\frac{E}{eV}\bigg)^{-2/7} 
\, , \quad E \, \geq \, 3 \times 10^{14} \; eV \, . 
\end{equation}
The decay of lepton pairs produced in the neutrino 
scattering process further enhances the spectrum of all neutrino 
species and allows neutrino emission, corresponding to present day 
energies $E \leq 10^{21}$ eV, to be detected from significantly 
higher redshifts than those given in (\ref{zn1}) and 
(\ref{zn2}) \cite{Yoshida}. The cascade enhancement due to decay 
is only relevant at high redshift where the neutrino mean free path is 
sufficiently small. In the case of electron and muon 
neutrinos, we will neglect the cascade enhancement of $z_{co_{\nu}}$ 
because the dominant term in the neutrino flux 
from VHS strings, (\ref{nueflux}) and (\ref{numuflux}), is proportional to 
$(1 - (z_{max_{\nu}} + 1)^{-1/2})$ and the modifications to the 
$\nu_e$, $\nu_{\mu}$, and $\nu_{\tau}$ spectra expected at Earth are 
negligible above $E \sim 10^{-3} m_X$ \cite{Yoshida}. 
The cascade enhancement at low energies increases as $m_X$ increases 
due to the greater energy in the initial jets.

\section[observations]{Comparison with Observations}
\label{sec:obs}

As we mentioned above, the diffuse fluxes of particles in the 
VHS scenario are produced directly by the decay of X-particles. 
The X-particles decay immediately after 
they are radiated from the long string segments. 
The result is that the particle fluxes are produced along the 
strings. 
The number of long strings per Hubble volume 
$\nu$ we noted in Section \ref{sec:intro} is approximately 13. As the 
Universe expands the comoving inter-string distance grows as 
\begin{equation}
\label{intersdist}
D_{ints} \propto \frac{2}{\sqrt{\nu}}(1 + z)^{-1/2} H_{o}^{-1} \; .
\end{equation}

\begin{figure}[ht]
\begin{center}
   \mbox{\epsfig{figure=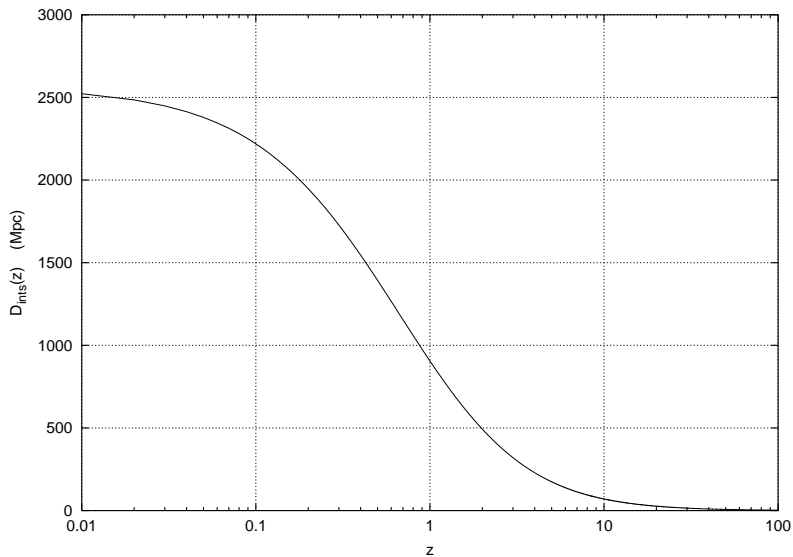,height=4.2in,angle=-90}}
\end{center}
\caption[...]{Average distance between long string segments as a 
function of the redshift.
\label{figavdlgstr}}
\end{figure}

Fig. \ref{figavdlgstr} depicts the evolution of the average distance 
between long string segments. The distance between the observer 
and a long string segment, $D_{ls}$ is estimated to be
\[
D_{ls} \sim  \frac{D_{ints}}{\sqrt[3]{2}} \; . 
\] 
If the distance today between the observer 
and the long string segment is larger than the maximum average 
propagation distance for particles 
arriving with an energy $E$ then the fluxes at Earth are exponentially 
suppressed \cite{Berez}. Based on Eq. (\ref{intersdist}) we see 
that presently the average distance to a long string segment is 
\begin{equation}
\label{intersdistnow}
D_{0,ls} \simeq 2000 \; \mbox{Mpc},
\end{equation}
which corresponds to a redshift $z_{ls} \simeq 0.58$.

The present average redshift to a long string segment $z_{ls}$ 
can be taken as a rough estimate of $z_{min}$, i.e., of the redshift 
corresponding to the latest time ($t_{istr}$) that the emission 
from the VHS string network can be considered isotropic. 
This sets an upper limit of the integral (\ref{flux}) 
and consequently in Eqs. (\ref{fluxsec}\textendash \ref{nutauflux}) 
we have  
\begin{equation}
\label{zmin}
z_{min} = z_{ls} \simeq 0.58 \; .
\end{equation} 
Hence only the particle fluxes emitted from the VHS network before 
$t_{istr}$ ($z > z_{min}$) are relevant for the diffuse flux.

A segment of long string could be substantially closer than $D_{0,ls}$, 
the expected average distance to the nearest string. In this case, 
a flux of cosmic rays protons (antiprotons) and photons above the GZK 
cutoff might be observed. 
However, these fluxes would be highly anisotropic in contradiction to the 
present observations of UHE events above the GZK cutoff \cite{agasanew}.

\subsection{Protons}
In what follows, we discuss the proton flux. Because the fragmentation 
functions and relevant interactions with the cosmic backgrounds are 
to first order charge symmetric, the expected antiproton flux from the 
VHS strings is the same as the proton flux.

The average maximum distance at the present time to the source of a 
proton arriving with energy $\sim 4 \times 10^{19}$ eV (the onset 
of the GZK cutoff) is $\alt 160$ Mpc (see Fig. \ref{figavdist}). 
This is equivalent to the source being at a redshift $z \alt 0.036$, 
much smaller than the distance from which the flux can be 
considered isotropic ($\simeq 2000$ Mpc). 
It implies that no isotropic flux of protons from 
VHS strings is expected at Earth above the GZK cutoff. 
On the other hand, a diffuse proton 
flux from the strings should reach the Earth at arrival energies 
$E \alt 6 \times 10^{18}$ eV, well below the UHE event energies.
Although no isotropic flux of protons at UHE energies is expected at 
Earth, the UHE protons  emitted at high redshifts will have 
interacted with the universal radio background 
(URB), CMB, and the infra-red/optical (IR/O) backgrounds 
to produce secondary fluxes of $\gamma$-rays, electrons, 
and neutrinos. These secondary fluxes could in principle  
be used to constrain the VHS model. However, in the context 
of the TD scenarios including the VHS scenario, 
these nucleon-induced secondary 
fluxes are negligible compared to the corresponding 
fluxes directly produced by the decay of the X-particles. 
This is because in the X-particle decay, far more direct 
$\gamma$-rays and neutrinos are produced than nucleons. 

\subsection{Photons and Charged Leptons}
The maximum redshift at 
which a photon arriving today with an energy $E$ could have been emitted 
is given in Fig. \ref{figphotabs}. If we compare this redshift to 
the present average distance to a long string ($z_{ls} \simeq 0.58$) 
we would deduce that the isotropic UHE $\gamma$-ray flux should 
also be exponentially suppressed. 
In the case of $\gamma$-rays, this is not the full story. 
Fig. \ref{figphotabs} describes the interaction length of the photons, 
i.e., the distance traveled by a UHE photon before it is 
absorbed in the cosmic backgrounds. 
The UHE photons initiate electromagnetic 
cascades as they travel in the cosmic backgrounds. 
An important consequence is that the effective penetration 
length of the cascade is considerably larger than the 
interaction length \cite{Bonom}.
In the VHS scenario, both the interaction length and the 
effective penetration length are much smaller than 
the distance to the nearest expected string (cf. Eq. (\ref{intersdist})).  
The result is that the $E \agt 10^{11}$ eV $\gamma$-ray are 
essentially subtracted from the arriving flux: The flux above 
$E \sim 10^{20}$ eV is depleted 
due to interactions with the URB; the flux in the range 
$10^{14} \alt E \alt 10^{20}$ eV is depleted by the CMB; and 
the flux in the range $10^{11} \alt E \alt 10^{14}$ eV 
is depleted by the IR/O backgrounds. 
The energy lost by the photons emitted with 
energies above $\sim 10^{11}$ eV at high redshifts 
is recycled into energies below $E \agt 10^{11}$ eV.
Analytic calculations \cite{ancascades} have shown that 
the electromagnetic cascade spectrum  has a generic shape 
below $ \agt 10^{13}$ eV, i.e., the secondary cascade 
radiation spectrum is insensitive to the injection spectrum. 
This implies that, even in the absence of the detection of 
direct UHE photons, the $\gamma$-ray flux below 
$E \agt 10^{11}$ eV can be used to constrain the VHS scenario. 
The charged leptons created in the X-particle decay 
also initiate electromagnetic cascades on the 
cosmic backgrounds. As in the case of the UHE photons, 
the energy of the UHE charged leptons is recycled and enhances the 
photon flux below $ \agt 10^{11}$ eV. 

The constraints on the electromagnetic emission in TD models with a 
spectral index $p=1$, which is applicable to the VHS scenario, 
stem from the limits on the present 
diffuse $\gamma$-ray background between $3 \times 10^{6}$ eV 
and $10^{11}$ eV \cite{egret}. These require 
that \cite{SJSB,ProtSta,SLBY,CA} 
\begin{equation}
\label{qem}
Q_{EM}^{0} \alt 3 \times 10^{-23} \; \mbox{eV} \;
\mbox{cm}^{-3} \; \mbox{sec}^{-1} \; , 
\end{equation}
for $h = 0.65$ where $Q_{EM}^{0}$ is the present rate of total 
energy injection into the electromagnetic channels. 

In the VHS scenario, we have from Eq. (\ref{jetdensity}) that 
the present rate of total energy injection into the electromagnetic 
channel (photon and charged lepton emission) is 
\begin{equation}
\label{Q0const}
Q_{EM-VHS}^{0} \sim 2.3 \times 10^{-12} \; \left[\frac{1}{2}\right] \;
\left[\frac{\nu}{13}\right] 
\left[G \mu \right] \left[\frac{h}{0.65}\right]^{3} \; \mbox{eV} \;
\mbox{cm}^{-3} \; \mbox{sec}^{-1} \; .
\end{equation}
Comparing to the $Q_{EM}^{0}$ limit in (\ref{qem}), we obtain a 
constraint on the energy density of a string in the VHS 
scenario of  
\begin{equation}
\label{photonconstr}
G \mu \alt 3 \times 10^{-11} \; .
\end{equation}
Here we have assumed an extragalactic magnetic 
field of $B_{exgal} \lesssim 10^{-11}$ G. Stronger values of the 
extragalactic magnetic field generate greater initial synchrotron 
losses at ultra-high energies, decreasing the UHE spectra, 
enhancing the cascade reprocessing into low energies, and 
thus lowering the upper limits on $Q_{EM}^{0}$ and $G \mu$. 
Because of the large distances to the sources, we have also neglected 
the uncertainties in the cosmic background URB and IR/O intensities 
\cite{CA}. 

\subsection{Neutrinos}
\begin{figure}[ht]
\begin{center}
   \mbox{\epsfig{figure=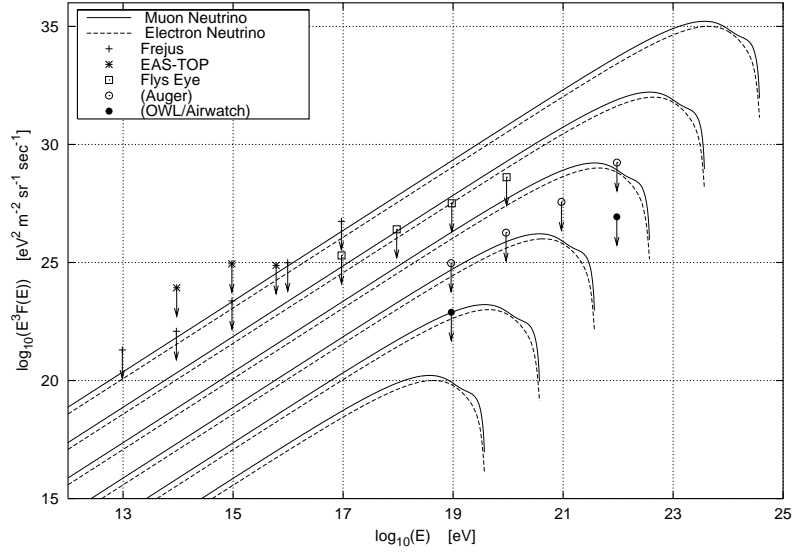,height=4.2in,angle=-90}}
\end{center}
\caption[...]{Electron and muon neutrino fluxes in the VHS cosmic string 
scenario (for $m_J = \eta / 2$) 
for various values of $G \mu$ (lines from top to bottom, 
$G \mu = 10^{-6}$, $10^{-8}, 10^{-10}, 10^{-12}, 10^{-14}, 10^{-16}$). 
The solid lines represent the $\nu_{\mu}$ flux and the dashed lines 
represent the $\nu_{e}$ flux.
Points with arrows represent upper limits on the diffuse neutrino 
flux from the EAS-TOP \cite{eastop}, the Fr\'ejus \cite{frejus} and the 
Fly's Eye \cite{eye} experiments. The upper limits stemming from the 
projected sensitivities for both the Pierre Auger project \cite{auger} 
and the OWL/Airwatch project \cite{owlairwatch} are also plotted.
\label{figemuflux}}
\end{figure}

\begin{figure}[ht]
\begin{center}
   \mbox{\epsfig{figure=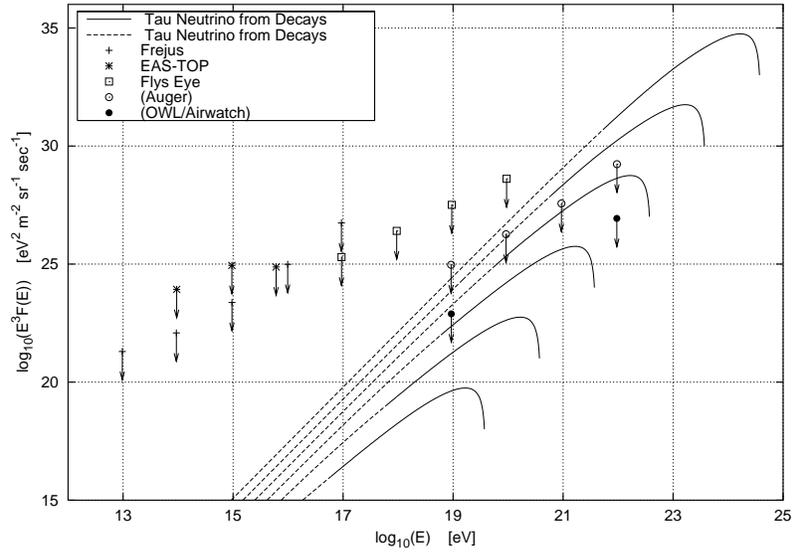,height=4.2in,angle=-90}}
\end{center}
\caption[...]{Tau Neutrino flux in the VHS cosmic string scenario 
(for $m_J = \eta / 2$) for 
various values of $G \mu$ (lines from top to bottom, 
$G \mu = 10^{-6}$, $10^{-8}, 10^{-10}, 10^{-12}, 10^{-14}, 10^{-16}$). 
For each value of $G \mu$, the range of energies where the $\nu_{\tau}$ 
from decays dominates the total $\nu_{\tau}$ flux is depicted by the 
solid section of the line. 
Points with arrows represent upper limits on the diffuse neutrino 
flux as in the previous figure.
\label{figtauflux}}
\end{figure}

The predicted flux of high energy neutrinos in the VHS 
scenario can be obtained from Eqs.(\ref{nueflux} - \ref{nutauflux}) 
using (\ref{zcutoff}), (\ref{zn1}), and (\ref{zn2}) to 
determine $z_{max_{\nu}}$. The resulting $\nu_{e}$ and $\nu_{\mu}$ 
fluxes are shown in Fig. \ref{figemuflux} and the $\nu_{\tau}$ flux 
in Fig. \ref{figtauflux}, for the  case when $m_J = \eta/2$ 
where $\eta$ is the scale of symmetry breaking. 

\begin{figure}[ht]
\begin{center}
   \mbox{\epsfig{figure=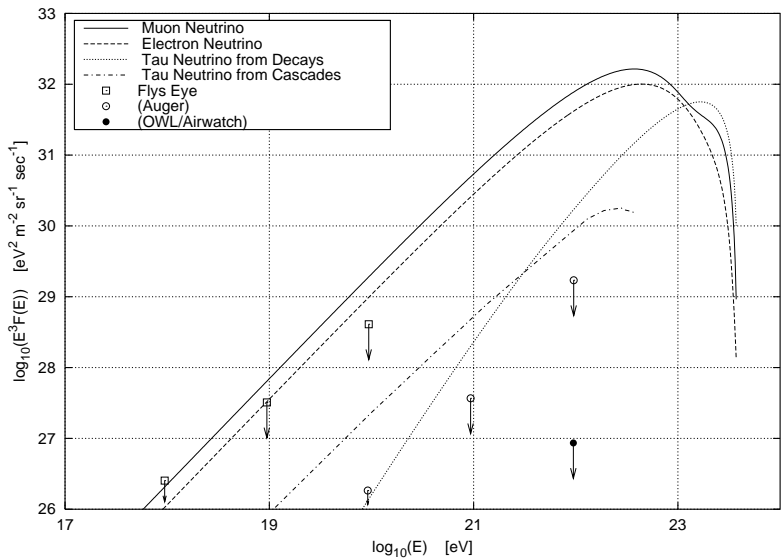,height=4.2in,angle=-90}}
\end{center}
\caption[...]{Neutrino flux in the VHS cosmic string scenario for 
various values of $G \mu = 10^{-8}$ 
The solid line represent the $\nu_{\mu}$ flux, the dashed lines 
represent the $\nu_{e}$ flux, and the dotted line the $\nu_{\tau}$ 
flux. The dash-dotted line represents the $\nu_{\tau}$ from the 
cascade of $\nu_{e}$ and $\nu_{\mu}$ due to the interactions with the 
relic neutrino background. 
Points with arrows represent upper limits on the diffuse neutrino 
flux as in the previous figures.
\label{fignuscasc}} 
\end{figure}

The cascading of the tau neutrinos off the relic neutrino 
backgrounds will have little effect on the tau neutrino spectrum 
shown in Fig. \ref{figtauflux}. However, a 
secondary $\nu_{\tau}$ component peaking at lower energies will 
be created in collisions of the string-produced $\nu_{e}$ and 
$\nu_{\mu}$ with the relic neutrino backgrounds. This 
$\nu_{e}$ and $\nu_{\mu}$ cascading should increase the net 
$\nu_{\tau}$ flux below $x \sim 10^{-3}$.
The $\nu_{\tau}$ signal expected at Earth from VHS strings will 
be the sum of these two components, as shown in Fig. \ref{fignuscasc} 
for $G \mu \sim 10^{-8}$ \cite{MWW,MW}. 
In Fig. \ref{fignuscasc} we have plotted the cascade component 
calculated in Ref. \cite{Yoshida} for a $p = 1$ TD scenario. 
This is only an estimate 
because the distance distribution of strings varies from model 
to model. We will address full modeling of the $\nu_{\tau}$ flux 
including cascades in a later paper. Note however that it is the high 
$x$ region, where the cascades have little effect, which is the most 
observationally interesting. 

Our new result is that the  $\nu_{e}$, $\nu_{\mu}$ and $\nu_{\tau}$ 
fluxes are of similar magnitude as $E \rightarrow m_J$, unlike all 
previous string emission calculations which predicted a negligible 
$\nu_{\tau}$ component at high $x$. In other string emission 
or X-decay scenarios, the $\nu_{e}$, $\nu_{\mu}$ and $\nu_{\tau}$ 
flux should also be of comparable magnitude as $x \rightarrow 1$. 
Because of the uncertainty in extrapolating to high emission energies, 
the precise ratio of the neutrino fluxes at high $x$ is not known
(also, as plotted in Fig. \ref{fignuscasc}, the extrapolation for 
$\nu_{\tau}$ is derived from HERWIG Monte Carlo simulations while the 
$\nu_{e}$ and $\nu_{\mu}$ functions are derived from the less rigorous 
approximation Eq. (\ref{primdecay}).) The actual ratio will be 
determined by the first step in the X-decay chain which is 
unknown and may be influenced by, for example, leptonic and 
SUSY decay channels (see Section \ref{sec:ext}).

The most recent observational limits come from the EAS-TOP \cite{eastop}, 
the Fr\'ejus \cite{frejus}, and the Fly's Eye \cite{eye} 
experiments. These experiments give upper limits on the 
cosmic ray neutrino flux in the energy range between $10^{13}$ eV 
and $10^{20}$ eV. In Figs. \ref{figemuflux} and \ref{figtauflux} we 
have plotted the observational limits using the updated 
$\sigma_{CC}(\nu N)$ cross-section \cite{Quigg}. 
By inspection, we see that the most stringent constraint comes from the 
highest energy $\nu_{\mu}$ flux in Fig. \ref{figemuflux} which gives 
an upper bound on $G \mu$ of 
\begin{equation} \label{presconst}
G \mu \, \alt \, 10^{-10} \, .
\end{equation}
This bound is derived at high $x$ so it should be little affected 
by the cascades which are neglected in Fig. \ref{figemuflux}.

We have also plotted on Figs. \ref{figemuflux} and 
\ref{figtauflux} the upper limits stemming 
from the planned sensitivities of the Pierre Auger project 
as estimated in Ref. \cite{auger}, and the OWL/Airwatch project as 
estimated in Ref. \cite{owlairwatch}. It 
is assumed that both experiments will collect data over a 
period of a few years. 
Their sensitivities span the energy range between $10^{19}$ eV 
and $10^{22}$ eV and would allow  more stringent upper limits on 
$G \mu$ up to 
\begin{equation}
G \mu \, \alt \, 10^{-12} \, ,
\end{equation}
in the case of null detection by the Pierre Auger project; and 
\begin{equation}
G \mu \, \alt \, 10^{-14} \, ,
\end{equation}
in the OWL/Airwatch case. Unlike the present constraint (\ref{presconst}), 
the Auger and OWL/Airwatch projects will constrain the $\nu_{\mu}$ flux 
in Fig. \ref{figemuflux} at energies $E \sim 10^{19}$ eV (i.e. the 
lower end of the energy range of these experiments).
The $\nu_{\tau}$ flux is below the present observational upper limits. 
The $\nu_{\tau}$ sensitivities of future experiments are expected to 
constrain $G \mu$, but these constraints 
are expected to be an order of magnitude weaker than those from the 
$\nu_{e}$ and $\nu_{\mu}$ fluxes. The slope of the $\nu_{\tau}$ 
spectrum is steeper than the slope of the observational upper limits 
for both present and future data (cf. Fig. \ref{figtauflux}). On the 
other hand, the $\nu_{\tau}$ component of a TD signature maybe easier 
to distinguish from other sources than the $\nu_{e}$ and 
$\nu_{\mu}$ components. 

\begin{figure}[ht]
\begin{center}
   \mbox{\epsfig{figure=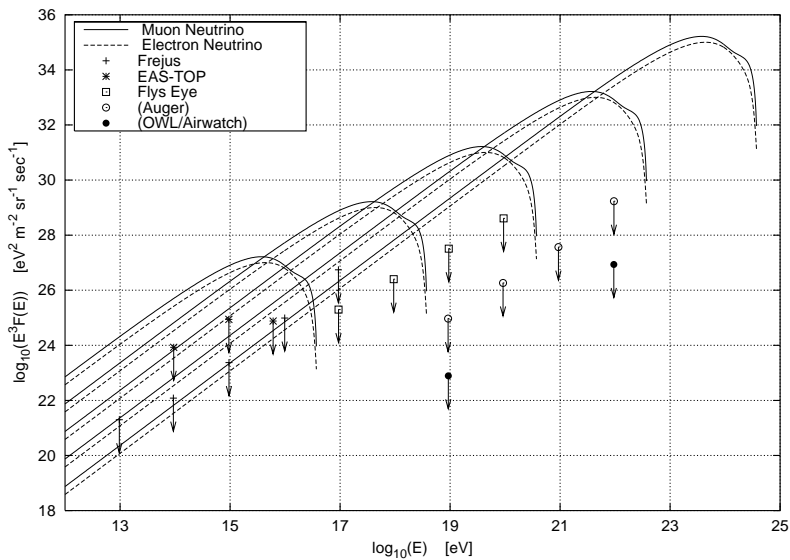,height=4.2in,angle=-90}}
\end{center}
\caption[...]{Electron and muon neutrino fluxes in the VHS cosmic 
string scenario for $G \mu = 10^{-6}$ and 
$m_J \sim 10^{25}, 10^{23}, 10^{21}, 10^{19}, 10^{17}$ eV  
(lines from bottom to top). The solid lines represent the 
$\nu_{\mu}$ flux and the dashed lines represent the $\nu_{e}$ flux.
Points with arrows represent upper limits on the diffuse neutrino 
flux as in the previous figures.
\label{fignuMj}}
\end{figure}

As is obvious from (\ref{asymptemu}), (\ref{asympttau}) and 
Fig. \ref{fignuMj}
the predicted flux of cosmic ray neutrinos increases as $m_J$ decreases. 
Thus, the VHS scenario with GUT-scale strings cannot be made compatible 
with observational constraints by lowering the energy scale $m_J$ of the 
initial jets (which likely happens if new physics evolves 
between the electroweak symmetry breaking scale and the GUT scale). 
Raising $m_J$ above $\eta$ would alleviate the conflict, but this 
appears quite unnatural.

The initial decay channels and branching ratios for the X-particle are 
unknown. The application of the extrapolated QCD-derived fragmentation 
functions to X-decay assumes implicitly that the branching ratios for 
$X \rightarrow  q_i + \bar{q_i}$ are the same as those for 
$e^{+}e^{-} \rightarrow  q_i + \bar{q_i}$. This may not be so for even 
hadronic decays at very high energies. The initial decay channel 
$X \rightarrow  q + l_X$, where $l_X$ is a charged lepton, should produce 
spectra which are decreased by no more than a factor of 2, compared with 
the $X \rightarrow  q + \bar{q}$ spectra (assuming equi-partition of the 
X-particle energy between the initial $q$ and $l_X$), with 
the exception of an $\nu_{\tau}$ spectrum enhancement as 
$E \rightarrow m_X/2$ due to decay of the initial $\tau$ \cite{MWW,MW}. 
The initial charged lepton $l_X$ would also generate EM cascades which 
are effectively attenuated at high energies in the VHS scenario, as we 
remarked in Section \ref{sec:jet} for the other EM decay products, 
but which would contribute to the spectra at low energies. This should 
increase the total energy going into the EM cascade channel by less 
than $\sim 50 \%$ (see Eq. (\ref{Q0const})). Pure leptonic decay 
channels $X \rightarrow  l_X + l'_X$, $X \rightarrow l_X + \nu_X$ or 
$X \rightarrow \nu_X + \nu'_X$ would produce substantially suppressed 
spectra, because of the low multiplicity per decay compared with hadronic 
jets, except for a possible neutrino spectra increase as 
$E \rightarrow m_X/2$ due to the initial step in the decay 
chain. In the case of purely leptonic decays, the nucleon and $\gamma$-ray 
spectra would be wholly created by the collisions of the emitted leptons 
with the cosmic ambient media. It has been shown that in TD models 
generally, pure leptonic decays can only be made interesting if the cosmic 
relic neutrinos are massive, non-relativistic and locally clustered 
thus enhancing the relative cascade-generated component (Ref. \cite{SLBY}). 
However, it is highly unlikely that the X-particles (being high energy 
gauge and scalar particles) would decay purely into leptons in the 
initial step. Because of the much higher multiplicity of hadronic jets, 
any scenario containing reasonable initial branching ratios for the 
X-particle (say, to within a factor of 2\textendash 3: $30\%$ pure 
hadronic, $30\%$ pure leptonic and $30\%$ hadronic and leptonic) would 
produce spectra dominated by the particles produced via the hadronic 
decay channels. 
Hence, even in the case of locally clustered relic neutrinos, the 
component produced in pure leptonic decays would be essentially 
dominated by the hadronic decays, with the possible exception of a UHE 
neutrino feature as $E \rightarrow m_X/2$.

\section[extensions]{Extensions to the Model}
\label{sec:ext}

In the analysis so far, we have assumed $m_{\nu}=0$. If neutrinos do 
possess a small mass, as present neutrino oscillation experiments hint, 
the predicted spectra from VHS strings are modified in two ways. 
Firstly, the interaction of the string-produced neutrinos with the 
$1.9^o$K relic background neutrinos is strongly enhanced at the resonance 
for $Z^{0}$ production \cite{Weiler}, $E_{res} = m_{Z}^{2}/2$ 
$m_{\nu} \sim 4 \times 10^{21}/(m_{\nu}/\mbox{eV})$ eV.  
(If neutrinos are massless, the $Z^{0}$ resonance occurs 
at $E_{res} \sim 3 \times 10^{24}$ eV \cite{Yoshida}.) 
The present upper limits from particle physics experiments on the 
neutrino masses are $m_{\nu_e} \alt 3$ eV, $m_{\nu_{\mu}} \alt 0.19$ 
MeV, and $m_{\nu_{\tau}} \alt 18.2$ MeV \cite{RPP}. 
The cosmological requirement that $\Omega_m \le 1$ places a stricter 
upper limit on the sum of the masses of the three neutrino 
species of $\alt 92 h^2$ eV \cite{Bludman}. Taken together these 
limits imply that at least one $Z^{0}$ resonance must occur for neutrinos 
today at some energy in the range $10^{20}$ eV 
$\alt E_{res} \alt 3 \times 10^{24}$ eV. The exact value(s) of 
$E_{res}$ will depend on whether or not the neutrinos have mass, and the 
value(s) of the neutrino mass(es) if non-zero. 
If the neutrino masses are non-degenerate, the $Z^{0}$ resonance will 
occur at two or three distinct $E_{res}$ in this range. In the case that 
$E_{res} < m_X$, the $Z^{0}$ resonance in the VHS scenario 
will produce a slight decrease in the predicted neutrino fluxes 
around $E_{res}$ and in the energy decade below $E_{res}$, 
and an accompanying slight enhancement on all spectra 
below $\sim 10^{-2} E_{res}$ due to secondary $Z^{0}$ decay 
products. The amplitudes of all spectra should be modified by 
less than a factor of 2. (See for example Ref. \cite{Yoshida} for the 
case of degenerate $m_{\nu} = 1$ eV neutrinos in other cosmic string 
models.) If the relic neutrinos are non-relativistic and clustered 
locally on a scale less than the attenuation length for photons and 
nucleons, the effect on the photon, proton and antiproton spectra 
may be enhanced \cite{YSL}.

Secondly, a non-zero neutrino mass may lead to oscillations between 
neutrino species as the flux propagates to Earth. In this case, the total 
number of neutrinos is preserved but the ratio of the flux in each 
species may depend on the neutrino energy, the distance traveled and 
neutrino masses. 
Recent observations by the SuperKamiokande, Kamiokande, MACRO and Soudan 
experiments of $\nu_{e}$ and $\nu_{\mu}$  neutrinos produced by 
cosmic rays colliding with the Earth's atmosphere strongly suggest that 
neutrino oscillations (at least between the $\nu_{\mu}$ and $\nu_{\tau}$ 
flavors) occur. [For a review of the current status of neutrino 
oscillation experiments and measurement implications see \cite{Akhm} 
and references therein]. The evidence for zenith-angle dependence in the 
SuperKamiokande data further strengthens this interpretation. Similarly, 
the long-standing observed deficit of solar neutrino flux may be 
resolved by $\nu_{e}$ oscillation 
into $\nu_{\mu}$ or $\nu_{\tau}$ although the solar neutrino experiments 
are not yet sufficiently consistent to favor a particular solution. 
The third set of evidence for neutrino oscillations comes from 
the LSND Collaboration accelerator experiment which has seen evidence for 
$\nu_{\mu}$ into $\nu_{e}$ oscillations. When combined with the 
atmospheric and solar data, this would require the existence of a 
fourth "sterile" neutrino species. The LSND measurements have yet to be 
independently confirmed. A number of new experiments are presently being 
commissioned with the objective of resolving these matters. The K2K, 
NuM-MINOS and CNGS experiments are expected to provide the definitive 
answer and exploration of the relevant parameter space. In these 
experiments, neutrino beams will be sent from accelerators to 
detectors over baselines of hundreds of kilometers. K2K is already 
taking data and the NuM-MINOS and CNGS will be operational in 2002 and 
2005, respectively.

Ignoring for now cosmological evolution, the probability of a 
relativistic neutrino undergoing vacuum flavor transition when 
propagating a distance $L$ is
\begin{eqnarray} \label{mu2tau}
P_{\nu_{\mu} \rightarrow \nu_{\tau}} =  sin^{2} (2 \,\theta_{0}) \, 
sin^{2}\Big(\frac{\Delta m_{atm}^2 \, L}{4 \hbar c E} \Big) & = &
sin^{2}( 2 \, \theta_{0}) \, 
sin^{2} \Biggl[  3.9 \times 10^{25}
\nonumber \\
& & {} \times
\Big(\frac{\Delta m_{atm}^2}{10^{-3} \ \mbox{eV}^{2}} \Big) 
\Big(\frac{L}{\mbox{Mpc}}\Big) 
\Big(\frac{E}{\mbox{eV}  }\Big)^{-1} \Biggr]
\end{eqnarray}        
in the 3-flavor solution. The atmospheric data to date are best fit by an 
amplitude of  $sin^2(\theta_0) = 0.82 - 1.0$ and a mass difference of 
$\Delta m_{atm}^2 = (2-6) \times 10^{-3} \mbox{eV}^2$ \cite{Akhm}. 
For models with distant cosmological sources (e.g. TDs and AGNs) 
and a neutrino ratio at emission of 
$F_{\nu_{e}} : F_{\nu_{\mu}} : F_{\nu_{\tau}} \simeq 1 : 2 : 0$ 
(in our case this occurs at $E << m_J$), the 
ratio of the spectra reaching the Earth will be 
$F_{\nu_{e}} : F_{\nu_{\mu}} : F_{\nu_{\tau}} = 1 : 1 : 1$ in the 
3-flavor model; similarly, in the 4-flavor model consistent with the 
LSND data, an initial emission ratio of 
$F_{\nu_{e}} : F_{\nu_{\mu}} : F_{\nu_{\tau}} : F_{\nu_{s}} \simeq 1 : 2 : 
0 : 0$ will become 
$F_{\nu_{e}} : F_{\nu_{\mu}} : F_{\nu_{\tau}} : F_{\nu_{s}} = 1 : 1 : 1 : 0$
 at Earth \cite{AHu}. Since we showed earlier that the current average 
distance to the nearest long string is 2000 Mpc, even the highest energy 
neutrinos emitted in the present era will have undergone transition before 
reaching Earth. In this case, at $E << m_J$, both the $\nu_{\mu}$ 
and $\nu_{\tau}$ flux reaching Earth should approximately equal half the 
$\nu_{\mu}$ flux expected without oscillations. In the case of the flux 
emitted in previous epochs, this will only be so provided the wavelength 
of the oscillation is much less than the mean free path of the $\nu_{\mu}$ 
and $\nu_{\tau}$ in the Universe. If the oscillation wavelength is greater 
than the mean free path then the cascade structure will be determined by 
the original neutrino flavors at emission. The mean free path 
decreases sharply as $z$ increases \cite{Yoshida}. For example, a 
$10^{23}$ eV neutrino at $z\sim 10^2$ has a mean free path of 
$\sim 100$ kpc \cite{Yoshida}. Since these too are the redshifts and 
energies at which cascading becomes relevant, cascade development 
should be predominantly determined by the neutrino flavors at emission. 
In the VHS scenario, however, the neutrino cascades have 
little influence on our regions of interest. Hence the predicted 
observable $\nu_{\mu}$ and $\nu_{\tau}$ VHS fluxes in the case with 
neutrino oscillations should be approximately half the unoscillated 
$\nu_{\mu}$ flux at $E << m_J$. As 
$E \rightarrow m_J/2$, i.e. in the region where the initial 
$\nu_{e}$, $\nu_{\mu}$, and $\nu_{\tau}$ VHS fluxes are of comparable 
magnitude, the ratio should also approach 
$F_{\nu_{e}} : F_{\nu_{\mu}} : F_{\nu_{\tau}} \simeq 1 : 1 : 1$ in the 
case with neutrino oscillations. The precise value of the modulated 
neutrino flux $F_{\nu_{i}}$ as $E \rightarrow m_J/2$ will depend on the 
initial absolute and relative string-produced fluxes which we regard 
as uncertain by more than a factor of 2. 

Additionally, one possible solution to the solar neutrino deficit 
problem is vacuum $\nu_{e} \rightarrow \nu_{\mu}$ or $\nu_{\tau}$ 
oscillations with $\Delta m_{\odot}^{2} \sim 10^{-10} \mbox{eV}^{2}$ 
and $sin^{2}2 \theta_{\odot} \sim 0.6 - 1.0$. In this case, the phase of 
the transition probability, 
$\phi_{\odot} = \Delta m_{\odot}^{2} L/(2 \hbar cE)$, 
is  $\alt 1$ for $E \agt 10^{21}$ eV neutrinos emitted 
today from a long string at 2000 Mpc. Thus the $\nu_{e}$ spectra at Earth 
would be determined by $\nu_{e} \rightarrow \nu_{\mu}$ or $\nu_{\tau}$ 
oscillation below $E \sim 10^{21}$ eV and exhibit negligible 
$\nu_{e}$ oscillation above $E \sim 10^{21}$ eV. If this particular solar 
deficit solution is combined with the solution implied by the atmospheric 
data $\Delta m_{atm}^{2} \sim 10^{-3} \; \mbox{eV}^{2}$, an initial ratio 
at emission of 
$F_{\nu_{e}} : F_{\nu_{\mu}} : F_{\nu_{\tau}} \simeq 1 : 2 : 0$ 
or $F_{\nu_{e}} : F_{\nu_{\mu}} : F_{\nu_{\tau}} \simeq 1 : 1 : 1$ 
would be modulated to 
$F_{\nu_{e}} : F_{\nu_{\mu}} : F_{\nu_{\tau}} \simeq 1 : 1 : 1$ 
reaching Earth because the wavelength for 
$\nu_{\mu} \rightarrow \nu_{\tau}$  
oscillation is much smaller than that of the oscillation solution 
associated with the atmospheric solar data. 
A detailed analysis of the detector signatures for various oscillation 
scenarios is presented in Refs. \cite{DRS1} and \cite{DRS2}.

Other extensions to the Standard Model could influence the predicted 
spectra. The existence of a supersymmetric (SUSY) sector would contribute 
extra decay channels in the fragmentation of the X-particle and jets. 
(See \cite{BK} and references therein for possible approaches to modeling 
SUSY channels for application to UHE decays). The net effect would be a 
skewing of the primary flux spectra to lower energies and a decrease in 
the energy reprocessed in high energy cascades. For example, in the p=1 
TD model studied in Ref. \cite{SLBY} with $m_X = 10^{16}$ GeV, 
including a SUSY sector decreases the spectra of all species expected 
at Earth by an order of magnitude at $E \agt 10^{-2} m_X$, shifts the UHE 
peaks in the spectra to lower energies by at least an order of magnitude 
and increases the amplitude of the $\nu_{e}$, $\nu_{\mu}$, $p$ and 
$\bar{p}$ spectra by about an order of magnitude at energies below the 
peaks. 
The predicted $E \alt 10^{15}$ eV photon spectrum,  
predominantly generated by EM cascades, though is unchanged 
because the same fraction of initial X-particle energy is 
reprocessed into low energies in the EM cascades with 
or without a SUSY sector in that model. In the case of 
the VHS scenario, these remarks must be convolved with the suppression of 
the UHE $\gamma$-ray and nucleon spectra due to the 
nearest string being farther than the relevant UHE attenuation lengths.

To date there is no experimental evidence for a SUSY sector and no unique 
theoretical predictions for the architecture of the SUSY sector and 
superparticle properties such as mass. Other extensions to the 
Standard Model, for example Technicolor or Leptoquarks, which also 
increase the number of degrees of particle freedom would be expected to 
similarly skew the spectra to lower energies. As well, extensions to the 
Standard Model may modify the high-energy cross-sections governing 
the interactions of the emission with the cosmic ambient media.

\section[concl]{Conclusions}
\label{sec:concl}
We have computed the ultra-high energy $\gamma-$ray, neutrino and 
cosmic ray fluxes in the VHS cosmic string scenario, in which the 
long string network loses its energy directly by particle emission. 
This is in contrast  to the standard cosmic string model with 
a scale-invariant distribution of string loops which lose energy 
predominantly by gravitational radiation. The predicted particle 
fluxes are much larger in the VHS scenario.

The predictions for the particle fluxes depend on extrapolating 
the QCD fragmentation function to energies much higher than current  
accelerator energies. New physics between the QCD scale and the scale 
$\eta = \sqrt{\mu}$ at which the defects are formed could 
significantly affect the final particle distributions. We 
parametrize this uncertainty by introducing a scale $m_J$ as 
the energy scale to which the QCD fragmentation functions can 
be extrapolated and regard $m_J$ as the initial jet energy. 
The smaller the value of $m_J$, the larger the number of jets 
which are generated by the initial emission from the cosmic 
string. We calculate the resulting particle fluxes as a function 
of both $m_J$ and $\mu$. A new aspect of our work is the 
computation of the significant tau neutrino fluxes directly produced 
by decay in the particle jets.

Our calculations show that the predicted flux $F(E)$ of 
$\gamma-$rays, electron and muon neutrinos and cosmic rays scales as 
$E^3 F(E) \sim E^{3/2}$, as in other defect models, 
whereas the present observations show a much weaker increase with $E$. 
Hence, the most stringent constraints on topological defect 
models come from the highest energies for which data exist.

Setting $m_J = \sqrt{\mu}/2$,  
the VHS strings with $G \mu \agt 10^{-10}$ produce electron 
and muon neutrinos in excess of the UHE observations. Thus, assuming 
that the strings evolve as in the VHS scenario, models with 
$G \mu \agt 10^{-10}$  or, equivalently a symmetry breaking scale of 
$\eta \agt 10^{23}$ eV, are ruled out by the UHE data. 
(This result is in accordance with the limit on $G \mu$ in 
Ref. \cite{Pijush98b}). 
At a given particle energy, the predicted fluxes reaching the Earth 
scale as $m_J^{-1/2}$ and increase with $G \mu$. Lowering $m_J$ 
increases the disagreement between predictions and observations, 
although the upper cutoff on the predicted spectra diminishes. 

A consistent but slightly stronger constraint on the VHS string 
scenario comes from the cascading of the electromagnetic emission 
off the cosmic radiation backgrounds and the potential conflict 
with the observed diffuse $3 \times 10^{9}$ eV $\alt E \alt$ 
$10^{11}$ eV gamma ray background as probed by EGRET. 
This requires $G \mu \lesssim 10^{-11}$. 

We conclude that, generically, GUT-scale strings are ruled out 
in the VHS scenario. However, VHS models with lighter 
$G \mu$, corresponding to lower symmetry-breaking scales, may be 
contributing interestingly to the observed UHE events. We reiterate 
that the VHS scenario is not universally accepted as giving 
the correct dynamics of cosmic strings. 

Finally, we note that the uncertainties inherent in the calculation of the 
expected fluxes in any TD model mean that the absolute and relative flux 
in each particle species should not be regarded as accurate to more 
than an order of magnitude in any model. The list of uncertainties 
includes: the mass and other properties of the X-particle(s); the form 
of the potential of the scalar field at string formation; the branching 
ratios and decay channels of the X-particle(s); the validity of the 
extrapolated fragmentation functions; the possibility of new 
particles (e.g. a SUSY sector) and other extensions to the Standard 
Model which may influence particle production, decay and propagation; 
the strength of the extragalactic magnetic field; the intensity of 
the cosmic IR background; and the Hubble constant. Taken together, 
these factors have the capability to increase or decrease the absolute 
and relative fluxes expected at Earth.

\begin{acknowledgments}
We would like to thank V. Berezinsky for his comments and suggestions.
This work has been supported (at Brown) in part by the US Department 
of Energy under contract DE-FG0291ER40688, Task A, and was performed 
in part while JHM held a NRC-NASA/JSC Senior Research Associateship.    
UFW has been supported at CENTRA-IST by ``Funda\c{c}\~ao para a 
Ci\^encia e a Tecnologia'' (FCT) under the program ``PRAXIS  XXI'' 
and in part by LIP-Lisbon.
\end{acknowledgments}

\appendix

\section*[app]{Neutrino Flux from Muon Decay}      
\label{sec:mdecay}
In the main text we computed the flux of particles 
resulting from two-body decay of charged pions in a jet. 
The muons which 
are produced in this decay are unstable and in turn decay via 
a three body decay process producing electrons, muon neutrinos 
and anti-electron neutrinos. 
In this Appendix we compute the full spectra neutrinos from the 
$\pi^{\pm} \rightarrow \mu^{\pm} \rightarrow \nu$ decay chain. 

Because of the finite muon rest mass, the muon and muon 
neutrinos created in the first stage of 
charged pion decay cannot carry an arbitrary 
fraction of the pion energy. In the limit $E_{\pi} \gg m_{\mu}$, 
the energies of the decay products lie between
\begin{equation} \label{range1}
E \, \in \, \bigl[ 0, \alpha^{-1} E_{\pi} \bigr] 
\end{equation}
for muon neutrinos and
\begin{equation} \label{range2}
E \, \in \, \bigl[ r E_{\pi}, E_{\pi} \bigr]
\end{equation}
for muons \cite{Stecker,Halzen}. Here, 
$r = \bigl(\frac{m_{\mu}}{m_{\pi}}\bigr)^2 = 1 - \alpha^{-1}$ where 
$\alpha$ is given by (\ref{alph}). The muon distribution is then 
\begin{equation}  \label{muondistr}
\frac{d \mu}{d E_{\mu}} = \alpha E_{\pi}^{-1} \; . 
\end{equation} 

In order to obtain the $\nu_{\mu}$-flux, we integrate 
(\ref{primdecay}) over the range (\ref{range1}), resulting 
in (\ref{secdecay}). To obtain the flux of muon decay products, 
we follow the method of Ref. \cite{Gaisser}. In the laboratory 
frame the distribution of leptons with energy 
$E = y E_{\mu}$ produced by the decay of a muon of energy 
$E_{\mu}$ is given by [Gaisser \cite{Gaisser} p. 92] 
\begin{equation}   \label{dn/dy}
\frac{d n}{d y} = g_0(y) - P_{\mu} g_1(y)
\end{equation}
in the limit $E_{\mu} \gg m_{\mu}, m_e$, where 
\[
g_0(y) = \frac{5}{3} - 3 y^2 + \frac{4}{3} y^3 
\]
\[
g_1(y) = \frac{1}{3} - 3 y^2 + \frac{8}{3} y^3
\]
for $\nu_{\mu}$ and 
\[
g_0(y) = 2 - 6 y^2 + 4 y^3 
\]
\[
g_1(y) = -2 + 12 y - 18 y^2 + 8 y^3
\]
for $\nu_e$. In (\ref{dn/dy}), $P_{\mu}$ is the projection of 
the muon spin in the muon rest 
frame along the direction of the muon velocity in the laboratory 
frame 
\[
P_{\mu} = \frac{2 E_{\pi} r}{E_{\mu} (1 - r)} - \frac{1 + r}{1 - r} \; . 
\]

Convolving the distribution (\ref{dn/dy}) with the distribution of 
muons from pion decay (\ref{muondistr}) and integrating over muon 
and pion energies, the neutrino spectrum in the laboratory frame is 
\begin{equation}    \label{ninte}
\frac{d N}{d E_{\nu}} = \int_{E_{\nu}}^{m_J} d E_{\pi} 
\int_{E_{min}}^{E_{\pi}} \frac{d E_{\mu}}{E_{\mu}} 
\frac{d N}{E_{\pi}} \frac{d \mu}{E_{\mu}} \frac{d \mu}{y} \; ,
\end{equation}
\[
E_{min} = min[r E_{\pi}, E_{\nu}] \; .
\]

For simplicity we will first evaluate (\ref{ninte}) for an initial 
pion distribution of the form 
\[
\frac{d N}{d E_{\pi}} = K E_{\pi}^{- a} \; , \hspace{2cm} a >0 \; . 
\]
Changing the order of integration, 
\[
\int_{E_{\nu}}^{m_J} d E_{\pi} \int_{E_{min}}^{E_{\pi}} d E_{\mu} 
\rightarrow \int_{E_{\nu}}^{m_J} d E_{\mu} 
\int_{E_{\mu}}^{E_{\mu}/r} d E_{\pi} \; ,
\]
the result is 
\begin{equation}  \label{ndistr}
\frac{d N}{d E_{\nu}} = K E_{\nu}^{- a} \frac{1-r^a}{a (1 - r)} 
\biggl[ f_0 + \frac{f_1}{1 - r} \Bigl[1 + r - \frac{2 a r}{a - 1} 
\Bigl( \frac{1 - r^{a - 1}}{1 - r^a} \Bigr) \Bigr] \biggr]
\end{equation}
where 
\[
f_0 = \frac{2 (a +5)}{a (a+2) (a +3)} - \frac{5}{3 a} 
\bigg(\frac{E_{\nu}}{m_J}\bigg)^{a} + \frac{3}{(a + 2)} 
\bigg(\frac{E_{\nu}}{m_J}\bigg)^{a + 2} - \frac{4}{3 (a + 3)} 
\bigg(\frac{E_{\nu}}{m_J}\bigg)^{a + 3}
\]
\[
f_1 = \frac{2 (1 - a)}{a (a + 2) (a + 3)} - \frac{1}{3 a} 
\bigg(\frac{E_{\nu}}{m_J}\bigg)^{a} + \frac{3}{(a + 2)} 
\bigg(\frac{E_{\nu}}{m_J}\bigg)^{a + 2} - \frac{8}{3 (a + 3)} 
\bigg(\frac{E_{\nu}}{m_J}\bigg)^{a + 3}
\]
for $\nu_{\mu}$ and 
\[
f_0 = \frac{12}{a (a+2) (a +3)} - \frac{2}{a} 
\bigg(\frac{E_{\nu}}{m_J}\bigg)^{a} + \frac{6}{(a + 2)} 
\bigg(\frac{E_{\nu}}{m_J}\bigg)^{a + 2} - \frac{4}{(a + 3)} 
\bigg(\frac{E_{\nu}}{m_J}\bigg)^{a + 3}
\]
\begin{eqnarray}
f_1 & = & \frac{12 (a - 1)}{a (a + 1) (a + 2) (a + 3)} + \frac{2}{a} 
\bigg(\frac{E_{\nu}}{m_J}\bigg)^{a} - \frac{12}{(a + 1)} 
\bigg(\frac{E_{\nu}}{m_J}\bigg)^{a + 1} 
\nonumber \\
& & {}+ \frac{18}{(a + 2)} 
\bigg(\frac{E_{\nu}}{m_J}\bigg)^{a + 2} - \frac{8}{(a + 3)} 
\bigg(\frac{E_{\nu}}{m_J}\bigg)^{a + 3}
\end{eqnarray}
for $\nu_e$. Because muons and antimuons are created with opposite 
spin polarization in charged pion decays, this spectrum also applies 
to the antiparticle decay chain. 

The distribution of pions produced in QCD jet decay can be 
approximated by the polynomial fragmentation function 
(\ref{secdecay}). Evaluating (\ref{ndistr}) for the appropriate 
values of $K$ and $a$ and extending the analysis to $a < 0$, we 
find the final distribution of neutrinos produced by jet decay to 
be approximately 
\begin{eqnarray}     
\frac{d N_{\nu_{\mu} + {\bar \nu_{\mu}}}}{d E_{\nu_{\mu}}} & \simeq &  
\frac{5}{8 m_J} \Bigg[ 12.48 
+ 0.44\;\biggl(\frac{E_{\nu_{\mu}}}{m_J} \biggr)^{-3/2} 
- 6.12\; \biggl(\frac{E_{\nu_{\mu}}}{m_J} \biggr)^{-1/2} 
\nonumber \\
& & {}- 7.16\; \biggl(\frac{E_{\nu_{\mu}}}{m_J} \biggr)^{1/2} \Bigg] \; , 
\qquad \quad E_{\nu_{\mu}} \le \frac{m_J}{\alpha} \; ,  \nonumber
\end{eqnarray}
where $\alpha$ is given in (\ref{alph}), for the muon neutrinos 
produced in the first stage of the pion decay, 
\begin{eqnarray}    
\frac{d N_{\nu_{\mu} + {\bar \nu_{\mu}}}}{d E_{\nu_{\mu}}}  & \simeq & 
\frac{5}{8 m_J} \Bigg[ 11.83 
+ 0.48\;\biggl(\frac{E_{\nu_{\mu}}}{m_J} \biggr)^{-3/2} 
- 5.80\; \biggl(\frac{E_{\nu_{\mu}}}{m_J} \biggr)^{-1/2} 
\nonumber \\
& & {}- 7.33\; \biggl(\frac{E_{\nu_{\mu}}}{m_J} \biggr)^{1/2} 
+ 0.97\; \biggl(\frac{E_{\nu_{\mu}}}{m_J} \biggr)^{2} 
\nonumber \\
& & {}- 0.15\; \biggl(\frac{E_{\nu_{\mu}}}{m_J} \biggr)^{3} 
\Bigg] \; , \qquad \quad E_{\nu_{\mu}} \le m_J \; , \nonumber
\end{eqnarray}
for the muon neutrinos produced by the decay of muons in the pion 
decay, and 
\begin{eqnarray}   
\frac{d N_{\nu_e + {\bar \nu_e}}}{d E_{\nu_e}} & \simeq &
\frac{5}{8 m_J} \Bigg[ 12.68 
+ 0.47\; \biggl(\frac{E_{\nu_{e}}}{m_J} \biggr)^{-3/2} 
- 5.96\;  \biggl(\frac{E_{\nu_{e}}}{m_J} \biggr)^{-1/2}
\nonumber \\
& & {}- 8.30\;  \biggl(\frac{E_{\nu_{e}}}{m_J} \biggr)^{1/2} 
- 0.11\; \biggl(\frac{E_{\nu_{\mu}}}{m_J} \biggr)^{1} 
\nonumber \\ 
& & {}+ 1.67\; \biggl(\frac{E_{\nu_{\mu}}}{m_J} \biggr)^{2} 
- 0.45\; \biggl(\frac{E_{\nu_{\mu}}}{m_J} \biggr)^{3} \Bigg] \; , \qquad 
\quad E_{\nu_{e}} \le m_J \; ,  \nonumber
\end{eqnarray}
for the final electron neutrinos.

\end{document}